\documentclass[traditabstract]{aa}
\pdfoutput=1
\usepackage{verbatim} 
\usepackage{amsmath}
\usepackage{amssymb}
\usepackage{graphicx}
\usepackage{booktabs}
\usepackage[authoryear]{natbib}
\usepackage[colorlinks=true,linkcolor=blue,urlcolor=blue,citecolor=blue]{hyperref}
\usepackage{xspace}
\usepackage{siunitx}

\makeatletter

\usepackage{txfonts}\usepackage{twoopt}
\bibpunct{(}{)}{;}{a}{}{,}
\usepackage{enumitem}

\graphicspath{{./}, {figures/}}

\newcommand{\msun}{{\rm M}_{\odot}}
\newcommand{\lsun}{{\rm L}_{\odot}}
\newcommand{\rsun}{{\rm R}_{\odot}}

\newcommand{\km}{{\rm km}}
\newcommand{\kms}{{\rm km\,s^{-1}}}

\newcommand{\bethe}{{\rm B}}
\newcommand{\mesa}{\mbox{\textsc{Mesa}}\xspace}
\newcommand{\eg}{e.g.\@\xspace}
\newcommand{\cf}{c.f.\@\xspace}
\newcommand{\ie}{i.e.\@\xspace}
\newcommand{\casea}{Case~A\@\xspace}
\newcommand{\caseb}{Case~B\@\xspace}
\newcommand{\casec}{Case~C\@\xspace}
\newcommand{\caseab}{Case~AB\@\xspace}
\newcommand{\casebb}{Case~BB\@\xspace}
\titlerunning{Pre-SN evolution, compact remnants and SN properties of stripped stars}
\authorrunning{F.R.N.~Schneider}
\makeatother
\begin{document}
\title{Pre-supernova evolution, compact object masses and explosion properties of stripped binary stars}
\author{%
F.R.N.~Schneider\inst{\ref{ZAH},\ref{HITS}}\thanks{fabian.schneider@uni-heidelberg.de}
\and Ph.~Podsiadlowski\inst{\ref{OXFORD}}
\and B.~M{\"u}ller\inst{\ref{MONASH}}
}
\institute{%
Astronomisches Rechen-Institut, Zentrum f{\"u}r Astronomie der Universit{\"a}t Heidelberg, M{\"o}nchhofstr.\ 12-14, 69120 Heidelberg, Germany\label{ZAH}
\and Heidelberger Institut f{\"u}r Theoretische Studien, Schloss-Wolfsbrunnenweg 35, 69118 Heidelberg, Germany\label{HITS}
\and Department of Physics, University of Oxford, Denys Wilkinson Building, Keble Road, Oxford OX1~3RH, United Kingdom\label{OXFORD}
\and School of Physics and Astronomy, Monash University, Clayton, Victoria 3800, Australia\label{MONASH}
}
\date{Received xxx / Accepted yyy}
\abstract{The era of large transient surveys, gravitational-wave observatories and multi-messenger astronomy has opened up new possibilities for our understanding of the evolution and final fate of massive stars. Most massive stars are born in binary or higher-order multiple systems and exchange mass with a companion star during their lives. In particular, the progenitors of a large fraction of compact object mergers, and Galactic neutron stars (NSs) and black holes (BHs) have been stripped off their envelopes by a binary companion. Here, we study the evolution of single and stripped binary stars up to core collapse with the stellar evolution code \mesa and their final fates with a parametric supernova (SN) model. We find that stripped binary stars can have systematically different pre-SN structures compared to genuine single stars and thus also different SN outcomes. The bases of these differences are already established by the end of core helium burning and are preserved up to core collapse. Consequently, we find that \casea \&~B stripped stars and single \& \casec stripped stars develop qualitatively similar pre-SN core structures. We find a non-monotonic pattern of NS and BH formation as a function of CO core mass that is different in single and stripped binary stars. In terms of initial masses, single stars of ${\gtrsim}\,35\,\msun$ all form BHs, while this transition is only at about $70\,\msun$ in stripped stars. On average, stripped stars give rise to lower NS and BH masses, higher explosion energies, higher kick velocities and higher nickel yields. Within a simplified population synthesis model, we show that our results lead to a significant reduction of the rates of BH-NS and BH-BH mergers with respect to typical assumptions made on NS and BH formation. Therefore, we predict lower detection rates of such merger events by, \eg, advanced LIGO than is often considered. We further show how certain features in the NS--BH mass distribution of single and stripped stars relate to the chirp-mass distribution of compact object mergers. Further implications of our findings are discussed with respect to the missing red-supergiant problem, a possible mass gap between NSs and BHs, X-ray binaries and observationally inferred nickel masses from Type Ib/c and IIP SNe.
}
\keywords{Gravitational waves -- binaries: general -- Stars: black holes -- Stars: massive -- Stars: neutron -- supernovae: general}
\maketitle
%
%
%
%
\section{Introduction}\label{sec:introduction}

The majority of massive stars (${\gtrsim}\,10\,\msun$) are born in binary or
higher-order multiple systems \citep[\eg][]{2007ApJ...670..747K,
2009AJ....137.3358M, 2012Sci...337..444S, 2013A&A...550A.107S,
2012MNRAS.424.1925C, 2014ApJS..213...34K} and a significant fraction of them
exchange mass with a companion during their lives \citep[${\gtrsim}\,70\%$; see
\eg][]{2012Sci...337..444S, 2017ApJS..230...15M}. This immediately implies that
a similar fraction of all supernovae (SNe) are from stars that experienced a
past binary mass-exchange episode. Mass exchange can proceed stably, but it may
also lead to unstable situations such that stars merge or evolve through a
so-called common-envelope phase \citep[see, \eg, reviews
by][]{2012ARA&A..50..107L, 2017PASA...34....1D}. In all cases, mass transfer
episodes can drastically change the further evolution of stars.

All known masses of Galactic neutron stars (NSs) and black holes (BHs) are
measured in close binary stars, \eg in X-ray binaries, double pulsar systems,
and pulsar and white dwarf binaries. These systems have in common that the first
born compact object went through a binary mass-transfer phase where the
progenitor star was stripped off its hydrogen-rich envelope, \ie they do not
originate from genuine single stars \citep[\eg][]{2003MNRAS.341..385P,
2006csxs.book..623T, 2014LRR....17....3P, 2017ApJ...846..170T,
2020A&A...638A..39L}. In some cases, also the second-born compact object formed
from a stripped star. It is yet unclear how envelope stripping has influenced
the masses of these NSs and BHs.

The situation is similar in compact object mergers that are now routinely
observed by gravitational-wave detectors: if the merging compact objects
originate from stars formed in the same binary system, they are also the
remnants of stars that lost their envelopes in a past binary mass-exchange
episode \citep[\eg][]{1998ApJ...506..780B, 1998A&A...332..173P,
2002ApJ...572..407B, 2003MNRAS.342.1169V, 2014A&A...564A.134M,
2016MNRAS.462.3302E, 2017NatCo...814906S, 2018MNRAS.480.2011G,
2018MNRAS.481.1908K}. Even for compact object mergers induced by dynamical
encounters in dense stellar systems (\eg star clusters), the likelihood that the
progenitor stars of the individual compact objects had a binary mass-exchange
history is non negligible as close binaries are also common in such
environments.

These aspects become more and more relevant in an era of gravitational-wave
astronomy and large transient surveys that deliver new insights into, \eg, the
NS--BH mass distribution, the supernova explosion physics and thereby massive
star evolution. To this end, we here study the evolution of single and stripped
binary stars up to the pre-SN stage and through core collapse to address two key
questions. First, how does envelope stripping by binary mass transfer affect the
pre-supernova structures of stars and, secondly, what are the consequences of
this for the ensuing core collapse and the outcomes of possible SN explosions?
Stripped stars lead to SNe of Type Ib/c, while stars with a hydrogen-rich
envelope at core collapse produce Type II SNe. But how do, \eg, the masses of
NSs and BHs, and the explosion energies differ in such cases? First steps in
trying to shed light on such questions have been taken
\citep[\eg][]{1996ApJ...457..834T, 1999A&A...350..148W, 2001NewA....6..457B,
2004ApJ...612.1044P, 2019ApJ...878...49W, 2020ApJ...890...51E}, but our
understanding remains incomplete. Here, we try to fill some of these gaps.

This paper is structured as follows. We describe our stellar evolution models
and how we model the core collapse and SN phase in Sect.~\ref{sec:methods}. Key
properties of the pre-SN structures of single and stripped stars are discussed
and compared in Sect.~\ref{sec:pre-sn-evolution-and-structure}. We then present
our findings on the consequences of the different pre-SN structures for the
explodability of stars, the compact object remnant masses, explosion energies,
kick velocities and nickel yields in Sect.~\ref{sec:core-collapse-and-outcome}.
In Sect.~\ref{sec:pop-syn}, we use our findings to study populations of compact
objects from stripped stars. This includes comparisons to Galactic compact
objects, compact object merger rates and chirp-mass distributions accessible
thanks to gravitational-wave astronomy. We discuss our results in
Sect.~\ref{sec:discussion} and conclude in Sect.~\ref{sec:conclusions}.

%
%
\section{Methods}\label{sec:methods}

We evolve massive, non-rotating stars with the
Modules-for-Experiments-in-Stellar-Astrophysics (\mesa) software package
\citep{2011ApJS..192....3P, 2013ApJS..208....4P, 2015ApJS..220...15P, 2018ApJS..234...34P, 2019ApJS..243...10P}
in revision 10398. All stars are evolved up to core collapse (defined as the
point when the iron-core infall velocity exceeds $950\,\kms$), and the pre-SN
structures are analysed with an extended version of the parametric SN code of
\citet{2016MNRAS.460..742M}. 

Two sets of stars are studied: stars ending their lives unperturbed as genuine
single stars and stars that lose their hydrogen-rich envelope at certain phases
in their evolution because of mass exchange with a binary companion. In
Sect.~\ref{sec:single-star-models}, we describe the models of the single stars
and in Sect.~\ref{sec:binary-star-models} those that are stripped off their
envelopes. We consider single stars of initial masses of $11\text{--}75\,\msun$
(in total 32 models) and stripped binary stars of $15\text{--}100\,\msun$ (in
total 102 models). Details on the parametric SN model are presented in
Sect.~\ref{sec:parametric-sn-code}.

\subsection{Single star models}\label{sec:single-star-models}

We consider non-rotating stars of solar metallicity ($Z=0.0142$) with a chemical
composition according to \citet{2009ARA&A..47..481A} and an initial helium mass
fraction of $Y=0.2703$. This chemical composition may be somewhat too metal poor
to properly represent our Sun as indicated by the solar modeling/composition
problem \citep[\eg][]{2009ApJ...705L.123S}. In our models, a higher metallicity
would primarily increase stellar wind mass loss and hence reduce the final
masses of stars at core collapse. Opacities are mainly from
\citet{1992ApJ...401..361R} and \citet{1996ApJ...464..943I}, supplemented by
low-temperature opacities of \citet{2005ApJ...623..585F} (see also the \mesa
instrument papers for more details). An approximate nuclear network consisting
of 21 base isotopes plus $^{56}\mathrm{Co}$ and $^{60}\mathrm{Cr}$ (\mesa's
\texttt{approx21\_cr60\_plus\_co56.net} network) is applied that covers all the
major burning phases of stars up to core collapse. Reaction rates are taken from
the JINA REACLIB database V2.2 \citep{2010ApJS..189..240C}. For example, the
${}^{12}\mathrm{C}(\alpha,\gamma){}^{16}\mathrm{O}$ reaction rate is from the
NACRE II compilation \citep{2013NuPhA.918...61X}. We use mixing-length theory
\citep{1965ApJ...142..841H} with a mixing-length parameter of
$\alpha_\mathrm{mlt}=1.8$.

To help numerically with the evolution of massive stars and their envelopes, we
enable \mesa's MLT++ that enhances convective-energy transport in low-density
envelopes of some stars and thereby suppresses envelope inflation. The Ledoux
criterion for convection is used, and we assume step convective-core
overshooting of 0.2 pressure-scale heights for core hydrogen and helium burning
\citep[\cf][]{2015A&A...575A.117S}. In all later nuclear burning phases and for
all convective shells, we switch off convective boundary mixing. Semi-convection
is applied with an efficiency factor of $\alpha_\mathrm{sc}=0.1$
\citep[\eg][]{2016ApJ...823..102C}. The latter efficiency is on the low side
with respect to the inferred interior mixing of stars in the Small Magellanic
Cloud \citep{2019A&A...625A.132S}. However, because of the overshooting in our
models, there are almost no semi-convective regions (\cf
Sect.~\ref{sec:single-vs-binary-star-evolution}) such that its exact mixing
efficiency is expected to be less relevant.

For the stellar winds, we essentially follow \mesa's ``Dutch'' wind scheme, but
modify the metallicity dependence \citep[see also][]{2006A&A...452..295E}. For
cool stars with effective temperatures $T_\mathrm{eff}<10,000\,\mathrm{K}$, we
apply the wind mass-loss rates, $\dot{M}$, of \citet{1990A&A...231..134N} with a
metallicity scaling of $\dot{M}\propto Z^{0.5}$ as suggested by
\citet{2011A&A...526A.156M} for red supergiants. For stars with effective
temperatures $T_\mathrm{eff}>11,000\,\mathrm{K}$, we use the
\citet{2000A&A...362..295V,2001A&A...369..574V} mass-loss prescription if the
surface hydrogen mass fraction $X_\mathrm{surf}>0.5$ and switch to Wolf--Rayet
(WR) wind mass loss if $X_\mathrm{surf}<0.4$. The WR wind mass-loss rates are
those of \citet{2000A&A...360..227N}, but with a metallicity scaling as
suggested by \citet{2005A&A...442..587V}. These scalings depend on the surface
chemical composition of nitrogen, carbon and oxygen, and are thought to
correspond to WN, WC and WO Wolf--Rayet stars. For surface hydrogen mass
fractions in the range $0.4\text{--}0.5$ and for effective temperatures in the
range $10,000\text{--}11,000\,\mathrm{K}$, we linearly interpolate between the
corresponding mass-loss rates. No extra mass loss is assumed for luminous-blue
variables (LBV) and other stars that might lose mass eruptively or
pulsationally.

\subsection{Binary star models}\label{sec:binary-star-models}

\begin{figure}
    \begin{centering}
    \includegraphics{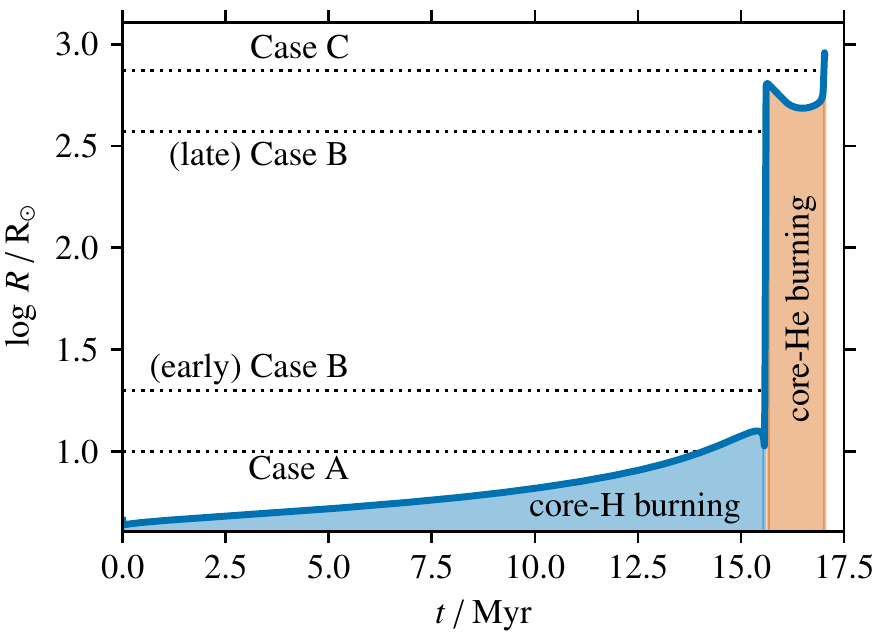} 
    \par\end{centering} \caption{Radius evolution of a $13\,\msun$ star and
    definition of \casea, B and~C mass transfer. In this work, we also
    distinguish between early and late \caseb, \ie mass transfer starting right
    after a star finishes its main-sequence evolution and just before core
    helium burning, respectively.}
    \label{fig:definition-mt-cases}
\end{figure}

Stars in a binary system can exchange mass once one star, the mass donor, has
grown to such a large radius that it overfills its Roche lobe. Depending on the
initial orbital separation of the two stars, mass exchange is initiated in
different evolutionary phases of the mass donor. In the initially closest
binaries, mass transfer starts when the donor star is still burning hydrogen in
its core (usually referred to as \casea mass transfer,
Fig.~\ref{fig:definition-mt-cases}). In initially wider systems, this only
occurs while the donor star crosses the Hertzsprung--Russell (HR) diagram after
finishing core hydrogen burning or climbs the giant branch (usually called
\caseb mass transfer). In initially even wider systems, mass transfer only
starts after the donor star finished core helium burning (\casec). The ensuing
mass-transfer episode can be stable or unstable, depending on the mass of the
donor, the orbital separation and the mass ratio of the binary star \citep[see,
\eg, figures~1--3 and 17--22 in][]{2015ApJ...805...20S}. Unstable mass transfer
leads to so-called common-envelope evolution or directly to a stellar merger. In
a successful common-envelope phase, most of the hydrogen-rich envelope of the
donor star is ejected while unsuccessful envelope ejections are thought to lead
to a merger of the two stars. In both cases, stable mass transfer and successful
common-envelope phases, the donor star loses almost its entire hydrogen envelope
\citep[how much hydrogen is left depends, \eg, on the metallicity of the star
and the yet uncertain wind mass-loss rates of stripped binary stars; see for
example][]{2017ApJ...840...10Y, 2017A&A...608A..11G, 2017A&A...607L...8V,
2019MNRAS.486.4451G, 2020A&A...637A...6L}.

Here, we do not want to follow the complex binary mass-exchange phases, but
focus on the effective outcome, namely the removal of the hydrogen-rich envelope
and the consequence of this for the further evolution of stars up to core
collapse. We remove the envelopes of stars in four evolutionary phases: towards
the end of the main sequence when the central hydrogen mass fraction falls below
0.05 (\casea), shortly after the star leaves the main sequence and starts to
cross the HR diagram (early \caseb, often corresponding to stable mass
transfer), shortly before the star ignites helium in its core (late \caseb,
often leading to unstable mass transfer that is thought to result in a
common-envelope phase) and after finishing core helium burning (\casec, often
unstable mass transfer that results in a common-envelope phase). Here we use the
usual definition of early and late \caseb mass transfer from donor stars with a
radiative and convective envelope, respectively. However, we do not find
systematic differences in early and late \caseb binaries and therefore discuss
them together as \caseb stripped stars in the rest of the paper. In
Fig.~\ref{fig:definition-mt-cases}, we indicate these four cases for a
$13\,\msun$ star. 

In our models, stars initially more massive than ${\approx}\,20\,\msun$ reach
their maximum radius before the onset of core helium burning and thus do not
experience \casec mass transfer. Below, we nevertheless consider \casec envelope
removal of such stars for academic purposes and clearly mark these stars in the
rest of this work. However, massive red supergiants are known to have extended
atmospheres \citep[\eg Betelgeuse,][]{2017A&A...602L..10O} and so-called wind
Roche-lobe overflow may increase the effective parameter space of \casec mass
transfer \citep[\eg][]{2007BaltA..16...26P, 2007ASPC..372..397M,
2013A&A...552A..26A, 2018A&A...618A..50S}.

In practical terms, we remove the envelopes on a timescale of 10\% of the
thermal timescales of stars. This is a good assumption for \caseb and~C mass
transfer given that \caseb mass transfer is driven by the thermal-timescale
expansion of the donor star and \casec mass transfer via common-envelope
evolution is thought to be near adiabatic, \ie even faster than what we assume.
Importantly, the chosen timescale for envelope removal is much faster than the
nuclear evolution of stars, \ie their cores do not evolve chemically during
envelope stripping. Applying an even faster timescale for envelope removal
therefore results in essentially the same pre-SN core structure (as is also
exemplified by the almost indistinguishable pre-SN core structures of stars
stripped in early and late \caseb mass transfer).

For \casea mass transfer, our assumption is less accurate and a
robust relation between the final stage after envelope removal and initial mass
is not possible within our simplified modelling. \casea mass transfer proceeds
via a fast, thermal-timescale mass-transfer phase followed by a slow, nuclear
timescale mass-transfer episode. During the latter and possibly subsequent
\caseab mass transfer, a significant fraction of the envelope may be removed
\citep[\eg][]{1967ZA.....65..251K, 1967AcA....17..355P, 2001A&A...369..939W}.
This sequence of envelope-removal episodes means that the core of the
mass-losing donor can further evolve while its envelope is continuously being
stripped off, and our simplified envelope-stripping model cannot reproduce this
evolution properly. We try to avoid part of these complications by restricting
our models to late \casea mass transfer. We stop the envelope removal once the
surface hydrogen mass fraction falls below 1\% for \caseb and~C mass transfer,
and once the surface hydrogen mass fraction is less than 10\% of that in the
core for late \casea mass transfer.

\subsection{Parametric supernova code}\label{sec:parametric-sn-code}

We compute explosion energies, compact remnant masses, nickel yields, and mean
kick velocities for our single and binary star models using the parametric
supernova model of \citet{2016MNRAS.460..742M}. Prior to shock revival, this
model estimates the neutrino heating conditions based on semi-empirical scaling
laws for the proto-NS radius, shock radius, the neutrino emission, and
the neutrino heating efficiency. If the model indicates that shock revival
occurs at some initial mass cut $M_\mathrm{i}$ before the gravitational neutron
star mass exceeds the stability limit (here assumed to be
$M_\mathrm{NS,grav}^\mathrm{max}=2.0\,\msun$, \ie a baryonic NS mass of about
$M_\mathrm{NS,by}^\mathrm{max}/\msun\approx M_\mathrm{NS,grav}^\mathrm{max}/\msun + 0.084
\cdot \left[M_\mathrm{NS,grav}^\mathrm{max}/\msun\right]^2 \approx2.336$; \cf
\citealt{1989ApJ...340..426L, 2001ApJ...550..426L}), the propagation of the
shock and the growth of the explosion energy are followed through a phase of
concurrent mass ejection and accretion. We account for energy input into the
explosion by neutrino heating and explosive burning, and for the binding energy
of matter swept up by the shock. Accretion and neutrino heating are assumed to
stop once the post-shock velocity reaches escape velocity at some mass
coordinate $M_\mathrm{f}$. If the explosion energy drops to zero at some point
after shock revival, or if accretion does not stop before the NS
exceeds the maximum mass of $2.0\,\msun$, we assume BH formation by
fallback. The original model of \citet{2016MNRAS.460..742M} does not predict the
amount of fallback, and we will therefore parametrically explore this.

\begin{figure*}
    \begin{centering}
    \includegraphics{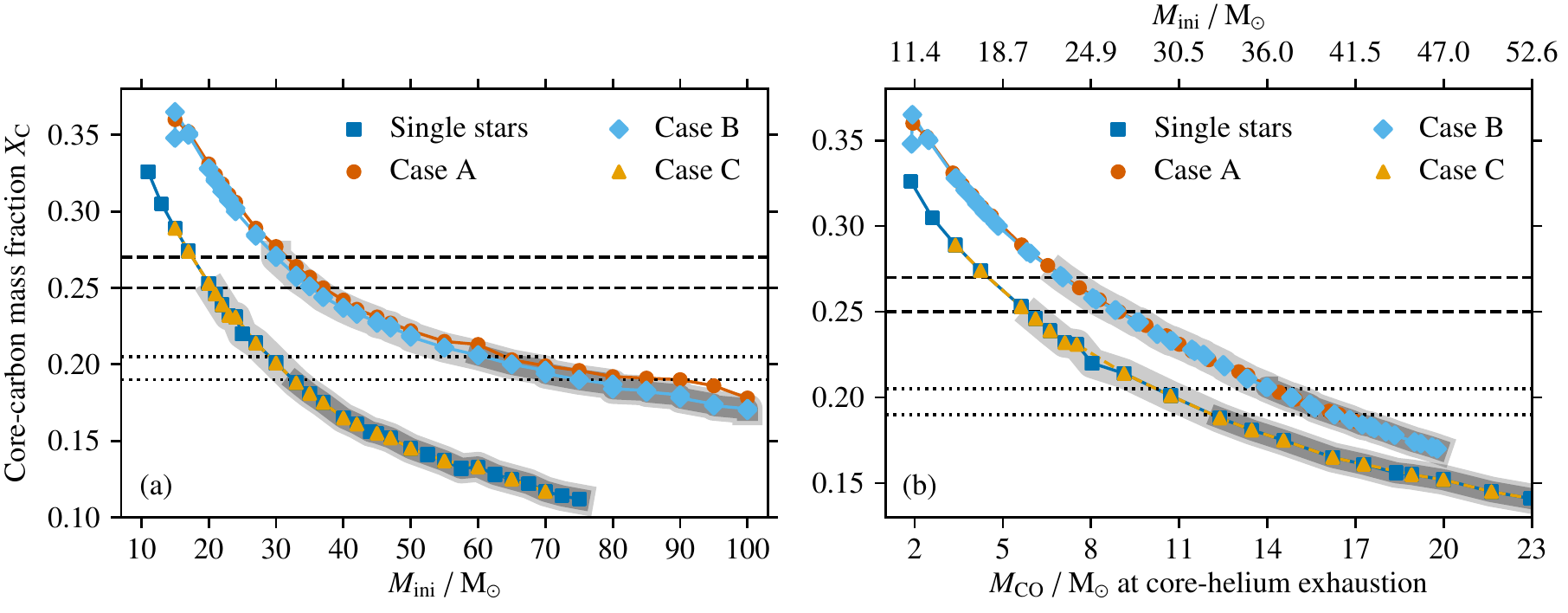} 
    \par\end{centering} \caption{Core carbon mass fraction $X_\mathrm{C}$ as a
    function of (a) initial mass $M_\mathrm{ini}$ and (b) CO core mass
    $M_\mathrm{CO}$ at the end of core helium burning. The initial masses
    $M_\mathrm{ini}$ of single stars corresponding to $M_\mathrm{CO}$ are
    provided at the top of panel b (\cf Fig.~\ref{fig:mini-mco}). The light-gray
    shading indicates radiative core carbon burning and the darkish-gray shading
    indicates radiative core neon burning. The dashed (dotted) lines show the
    core carbon mass fractions at the end of core helium burning below which our
    models burn carbon (neon) in their cores radiatively. In panel (b), we zoom
    in on the two branches and the ranges in $X_\mathrm{C}$ and $M_\mathrm{CO}$
    do not correspond directly to those in panel (a).}
    \label{fig:core-c-mass-fraction}
\end{figure*}

The mass $\Delta M=M_\mathrm{f}-M_\mathrm{i}$ and kinetic energy
$E_\mathrm{kin}$ (which is approximately equal to the explosion energy
$E_\mathrm{expl}$) of the post-shock matter at the ``freeze-out'' of accretion
determine the NS momentum $p_\mathrm{NS}=\alpha \sqrt{2\Delta M \,
E_\mathrm{kin}}$, where the parameter $\alpha$ characterizes the asymmetry of
the neutrino-heated ejecta \citep{2017ApJ...837...84J, 2018MNRAS.481.4009V}.
Although the degree of asymmetry is not universal \citep{2017ApJ...837...84J}, a
case can be made for a modest scatter in $\alpha$ due to the dominance of
unipolar explosions \citep{2018ApJ...856...18K, 2019MNRAS.484.3307M}, perhaps
with a tail towards low $\alpha$ from explosions with bipolar geometry
\citep{2008A&A...477..931S, 2019MNRAS.484.3307M}. For simplicity, we assume a
constant asymmetry parameter as in \citet{2018MNRAS.481.4009V} and compute the
mean NS kick velocity $v_\mathrm{kick}$ as
\begin{equation}
    v_\mathrm{kick} = 0.195 \frac{\sqrt{2 \Delta M E_\mathrm{expl}}}{M_\mathrm{rm,grav}},
    \label{eq:vkick}
\end{equation}
where $M_\mathrm{rm,grav}$ is the gravitational mass of the compact remnant
(either a NS or BH). The factor of 0.195 is a calibration factor chosen to match
the best-fit $\sigma$ of \citet{2005MNRAS.360..974H} of $\sigma\approx265\,\kms$
(see Sect.~\ref{sec:kicks}). We note that the kick mechanism will result in kick
velocities around the above computed mean value (Eq.~\ref{eq:vkick}) with some
dispersion. Here, we only consider the mean value and distributions thereof. In
Appendix~\ref{sec:relation-SN-code-structure}, we show how $M_\mathrm{i}$ and
$\Delta M$ are closely related to the two structural parameters of pre-SN models
introduced by \citet{2016ApJ...818..124E} to classify the explodability of
stars.

We use a different calibration of the SN engine than in the original work of
\citet{2016MNRAS.460..742M}. This is to obtain an average explosion energy of
Type IIP SNe in the range $0.5\text{--}1.0\,\bethe$ (\eg
\citealt{2009ApJ...703.2205K} suggest $0.9\,\bethe$). The average explosion
energy of Type IIP SNe is $0.69\pm0.17\,\bethe$
(Sect.~\ref{sec:explosion-energy-and-nickel-mass}) for our chosen calibration
parameters of the shock compression-ratio, $\beta=3.3$, and the shock expansion
due to turbulent stresses, $\alpha_\mathrm{turb}=1.22$, \citep[original values
are $\beta=4.0$ and $\alpha_\mathrm{turb}=1.18$;][]{2016MNRAS.460..742M}.

%
%
\section{Pre-SN evolution and structure}\label{sec:pre-sn-evolution-and-structure}

\subsection{Single vs.\ stripped binary stars towards core collapse}\label{sec:single-vs-binary-star-evolution}

\begin{figure}
    \begin{centering}
    \includegraphics{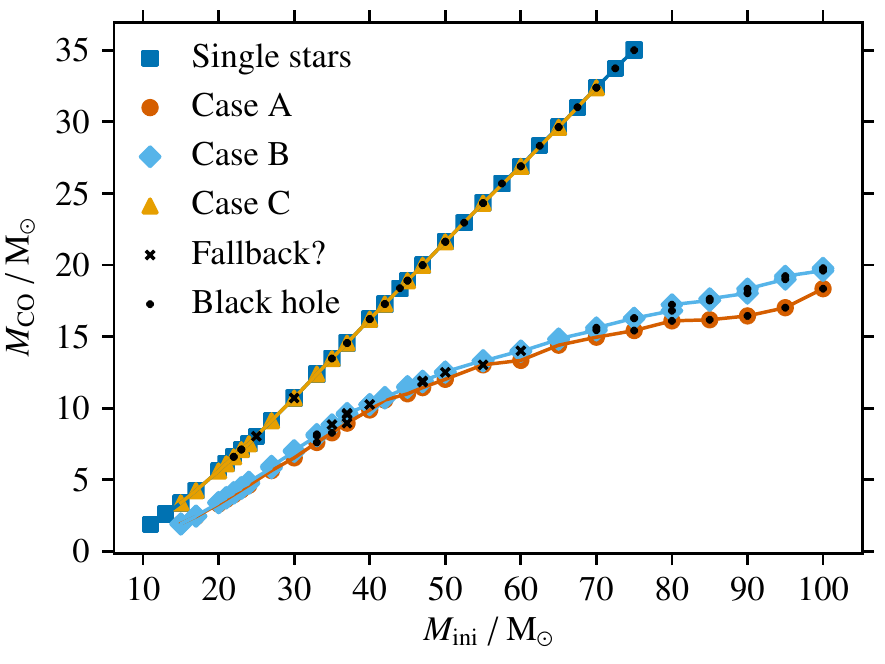} 
    \par\end{centering} \caption{CO core masses $M_\mathrm{CO}$ at the
    core-helium exhaustion as a function of initial mass $M_\mathrm{ini}$ of our
    single stars and the \casea, B and~C stripped stars. Models leading to BHs
    and likely experiencing significant fallback are indicated (see
    Sects.~\ref{sec:compact-remnant-masses} and~\ref{sec:ns-bh-mass-distr} for
    more details).}
    \label{fig:mini-mco}
\end{figure}

In massive single stars after core helium burning, the evolution of the envelope
and the core decouple after the core contracted and heated up sufficiently:
there is a steep drop in pressure at the hydrogen-helium interface, where also
the hydrogen-burning shell is located. The core no longer ``feels'' much of the
hydrogen-rich envelope (\eg in our initially $17\,\msun$ single star at core
helium exhaustion, the pressure drops from a mass coordinate of
${\approx}\,6\,\msun$ by more than ten orders of magnitude over
$1\text{--}2\,\msun$). The further core evolution and advanced nuclear-burning
phases of (non-rotating) massive stars therefore essentially depend on the
helium core mass $M_\mathrm{He}$, the CO core mass $M_\mathrm{CO}$ and the
amount of carbon left by helium burning $X_\mathrm{C}$. Especially
$X_\mathrm{C}$ and $M_\mathrm{CO}$ are thought to be most relevant, to the
extent that the evolution beyond helium burning is sometimes considered
bi-parametric \citep[\eg][]{2020ApJ...890...43C, 2020arXiv200503055P}. Single
stars populate a well-defined sequence in this $M_\mathrm{CO}$--$X_\mathrm{C}$
plane (Fig.~\ref{fig:core-c-mass-fraction}) as well as in the
$M_\mathrm{ini}$--$X_\mathrm{C}$ plane because of the one-to-one relation of
$M_\mathrm{CO}$ and $M_\mathrm{ini}$ (Fig.~\ref{fig:mini-mco}). 

The carbon abundance determines the strength of the ensuing carbon and later
burning stages \citep[\eg][]{1996ApJ...463..297B, 2001NewA....6..457B,
2004ApJ...612.1044P, 2020ApJ...890...43C, 2020MNRAS.492.2578S}. For example in
our single stars, core carbon burning turns radiative for a carbon mass fraction
$X_\mathrm{C}\,{\lesssim}\,0.25$ at core helium exhaustion and also neon burning
turns radiative for $X_\mathrm{C}\,{\lesssim}\,0.19$. With a lower mass fraction
of carbon and neon, the energy generated by carbon and neon burning can be
mostly transported away by neutrinos such that no core convection
develops\footnote{Oxygen burning and beyond proceed under convective
conditions.}. We will see below that radiative carbon and neon burning are
indicators for the strength of advanced nuclear burning in general and hence the
compactness of cores at core collapse.

Stars that lost their hydrogen-rich envelopes in \casea and~B mass transfer
reach genuinely different conditions in terms of $X_\mathrm{C}$ and
$M_\mathrm{CO}$ at the end of core helium burning than single stars
(Fig.~\ref{fig:core-c-mass-fraction}). Consequently, their core evolution
towards supernova is also inherently different from that of single stars
\citep[see \eg][]{1996ApJ...457..834T, 1996ApJ...463..297B, 2001NewA....6..457B,
1999A&A...350..148W, 2004ApJ...612.1044P, 2019ApJ...878...49W}. By definition,
\casec mass transfer leads to the same core conditions as in single stars at the
end of core helium burning. However, this does not also automatically imply that
the interior structure at core collapse is the same (see below). Also, the
mapping from $M_\mathrm{CO}$ to initial mass differs greatly in \casea and~B
stripped stars compared to single and \casec stripped stars
(Fig.~\ref{fig:mini-mco}). This is particularly important when considering
populations of stars as we do later (\cf Fig.~\ref{fig:core-c-mass-fraction}a).

We find that the early and late \caseb models populate the same branches in the
$M_\mathrm{CO}$--$X_\mathrm{C}$ plane (Fig.~\ref{fig:core-c-mass-fraction}) and
also their pre-SN structures are practically indistinguishable. As outlined in
Sect.~\ref{sec:binary-star-models}, nuclear burning from our early to late
\caseb systems does not advance significantly during the star's
thermal-timescale evolution. Analogously, our \casea and \caseb models are also
quite similar (\eg in the $M_\mathrm{CO}$--$X_\mathrm{C}$ plane), because we
only consider late \casea mass transfer where the donor star's interior is not
too dissimilar to that of early \caseb donor stars. Note, however, that our
\casea models are limited in properly representing the distinct mass transfer
episodes usually expected in \casea mass transfer
(Sect.~\ref{sec:binary-star-models}). Properly treating \casea mass transfer and
considering earlier \casea systems than studied here is required to establish
whether such stripped stars can yet populate another region in the
$M_\mathrm{CO}$--$X_\mathrm{C}$ plane and thus give rise to other pre-SN
structures.

In our models, the two branches in the $M_\mathrm{CO}$--$X_\mathrm{C}$ plane
clearly persist up to core neon exhaustion and are still visible by different
chemical compositions after core silicon burning. Because of this, stripped
stars are expected to produce systematically different chemical yields than
single stars, as will be studied in a forthcoming publication. 

\begin{figure*}
    \begin{centering}
    \includegraphics{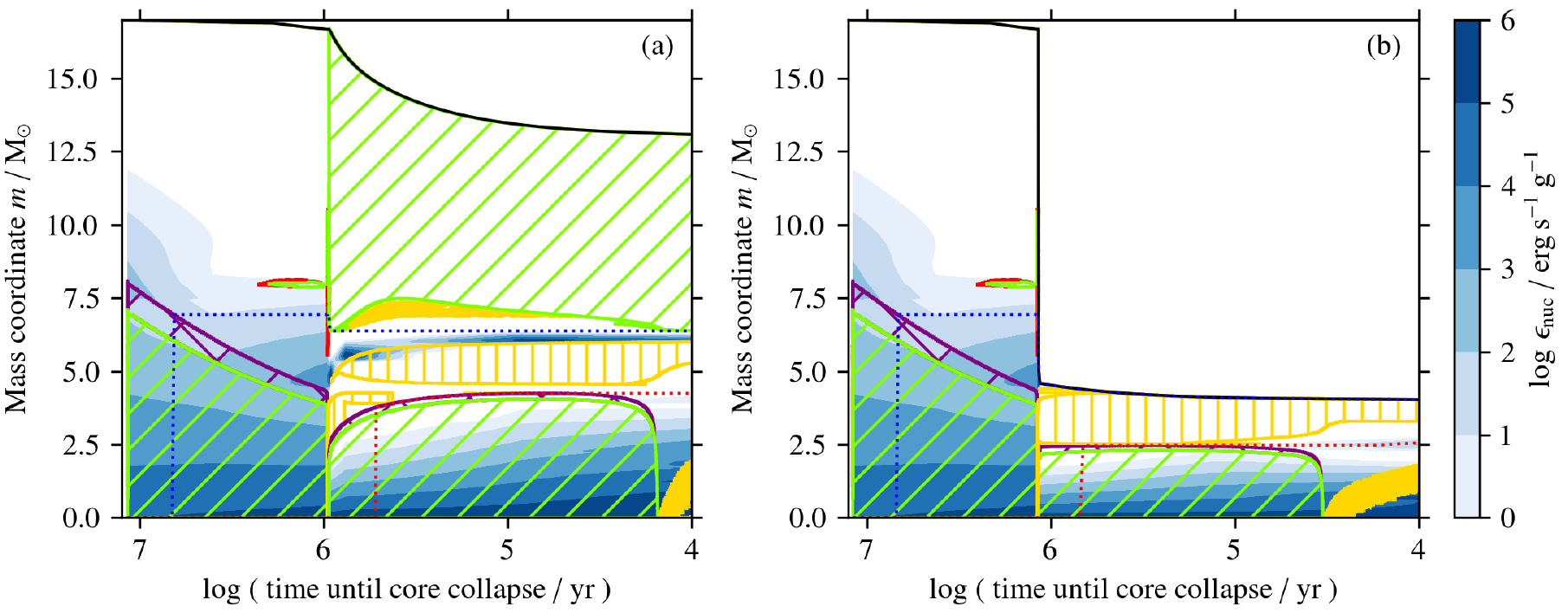} 
    \par\end{centering} \caption{Kippenhahn diagrams of core hydrogen and core
    helium burning of stars with an initial mass of $17\,\msun$. The evolution
    of a genuine single star (left panel a) is contrasted with that of a star
    that underwent (late) \caseb mass transfer (right panel b). The blue
    color-coding shows energy production by nuclear burning, and the green,
    yellow, purple and red hatched regions denote convection, thermohaline
    mixing, convective overshooting and semi-convection, respectively. The blue
    and red dotted lines indicate approximate helium and carbon cores, here
    defined as that mass coordinate where the helium and carbon mass fractions
    first exceed 0.5.}
    \label{fig:kipp-comparison}
\end{figure*}

The two branches in Fig.~\ref{fig:core-c-mass-fraction} can be understood as
follows \citep[see also][]{1996ApJ...463..297B, 2001NewA....6..457B}. During
core helium burning, helium nuclei are first burnt into carbon and only at a
later time is carbon converted into oxygen via the
${}^{12}\mathrm{C}(\alpha,\gamma){}^{16}\mathrm{O}$ reaction. In single stars,
the convective, helium-burning core grows over time thanks to a hydrogen-burning
shell that adds mass to the helium core (Fig.~\ref{fig:kipp-comparison}a). In
contrast, the mass of the convective, helium-burning core of stars that have
lost their hydrogen-rich envelope (and hence do not have a hydrogen-burning
shell) stays roughly the same or even decreases\footnote{Whether the
convective-core size stays constant or decreases depends on the mass loss of a
star during core helium burning. In massive stars with strong winds, the mass of
the convective core decreases.} during core helium burning
(Fig.~\ref{fig:kipp-comparison}b). Consequently, there are fewer
$\alpha$-particles available to convert carbon into oxygen such that these stars
have larger carbon abundances at the end of core helium burning.

There are two further aspects that contribute to the two branches in
Fig.~\ref{fig:core-c-mass-fraction} and to differences between stripped and
single stars at core collapse. First, stars that have lost their hydrogen-rich
envelopes are compact WR stars that lose mass at different rates than extended
(super)giants. In particular, the winds directly decrease the mass of the helium
cores. For example, this increases the carbon abundance at the end of core
helium burning\footnote{Switching off WR winds would reduce the separation
between the two branches in Fig.~\ref{fig:core-c-mass-fraction} and turning up
these wind mass-loss rates would increase it.} and may affect the central
temperature and density. Moreover, this also applies to \casec stripped stars
such that these models tend to have smaller helium core masses at core collapse
than their single star counterparts, which can induce differences in the pre-SN
structure despite both models falling onto the same branch in
Fig.~\ref{fig:core-c-mass-fraction} at the end of core helium burning. Secondly,
the missing weight of hydrogen-rich envelopes reduces the pressure near the
surface of stripped stars, so their cores tend to evolve similar to those of
single stars with a slightly less-massive helium core.

\begin{figure*}
    \begin{centering}
    \includegraphics{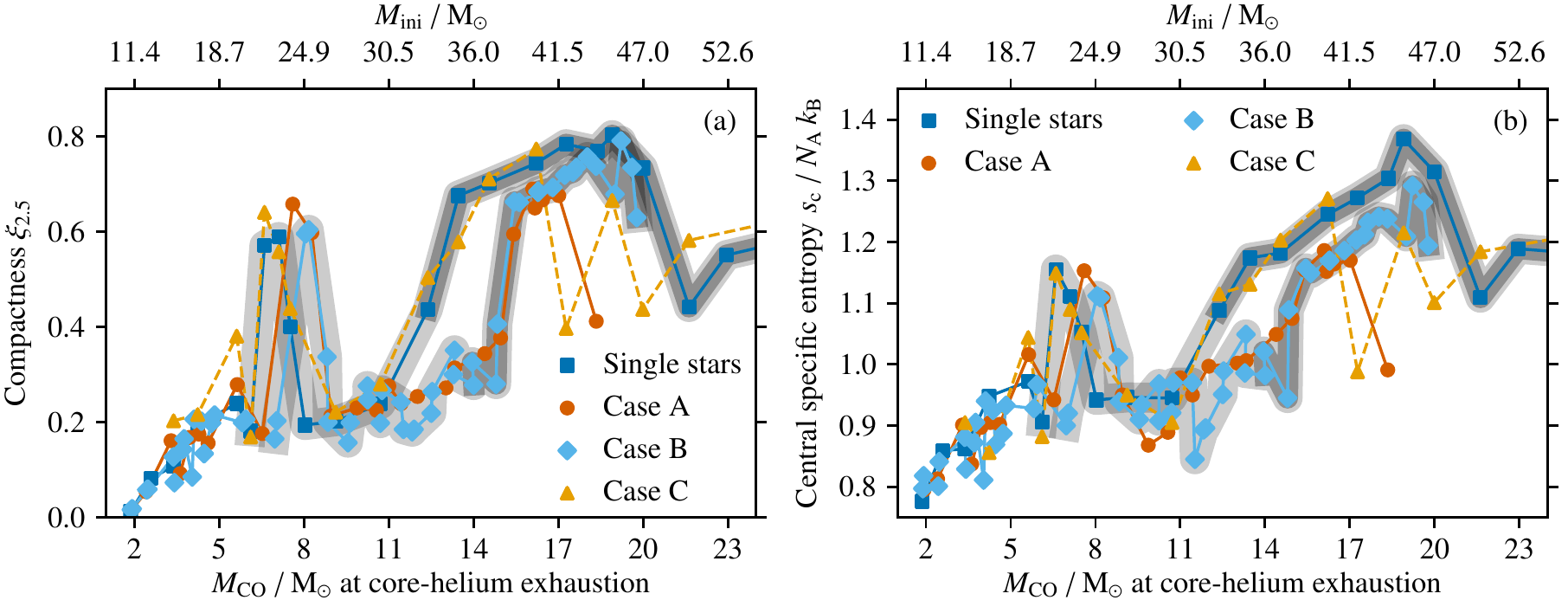} 
    \par\end{centering} \caption{Compactness $\xi_{2.5}$ (left panel a) and
    dimensionless central specific entropy $s_\mathrm{c}$ (right panel b) at
    core collapse as a function of CO core mass $M_\mathrm{CO}$. As in
    Fig.~\ref{fig:core-c-mass-fraction}, initial masses $M_\mathrm{ini}$
    corresponding to the CO core masses of single stars are shown at the top
    (\cf Fig.~\ref{fig:mini-mco}), and the light-gray and darker-gray shadings
    are for radiative core carbon and core neon burning, respectively.}
    \label{fig:compactness-and-central-entropy}
\end{figure*}

Because of these differences, also the conditions for which carbon and neon
burning turn radiative change in \casea and~B stripped stars (in \casec models
we could not find differences with respect to single stars). Carbon burning
proceeds under radiative conditions for $X_\mathrm{C}\,{\lesssim}\,0.270$ and
neon burning for $X_\mathrm{C}\,{\lesssim}\,0.205$ at the end of core helium
burning, while these limits are $X_\mathrm{C}\,{\lesssim}\,0.250$ and
$X_\mathrm{C}\,{\lesssim}\,0.190$ in our single and \casec stripped stars. It is
also evident from our models that the carbon abundance alone does not determine
whether the ensuing carbon- and neon-burning phases are convective or radiative,
but it is a combination of $X_\mathrm{C}$ and $M_\mathrm{CO}$ \citep[see
also][]{2020MNRAS.492.2578S}.

\subsection{Pre-SN compactness, central entropy and iron core mass}\label{sec:pre-sn-structure}

\begin{figure}
    \begin{centering}
    \includegraphics{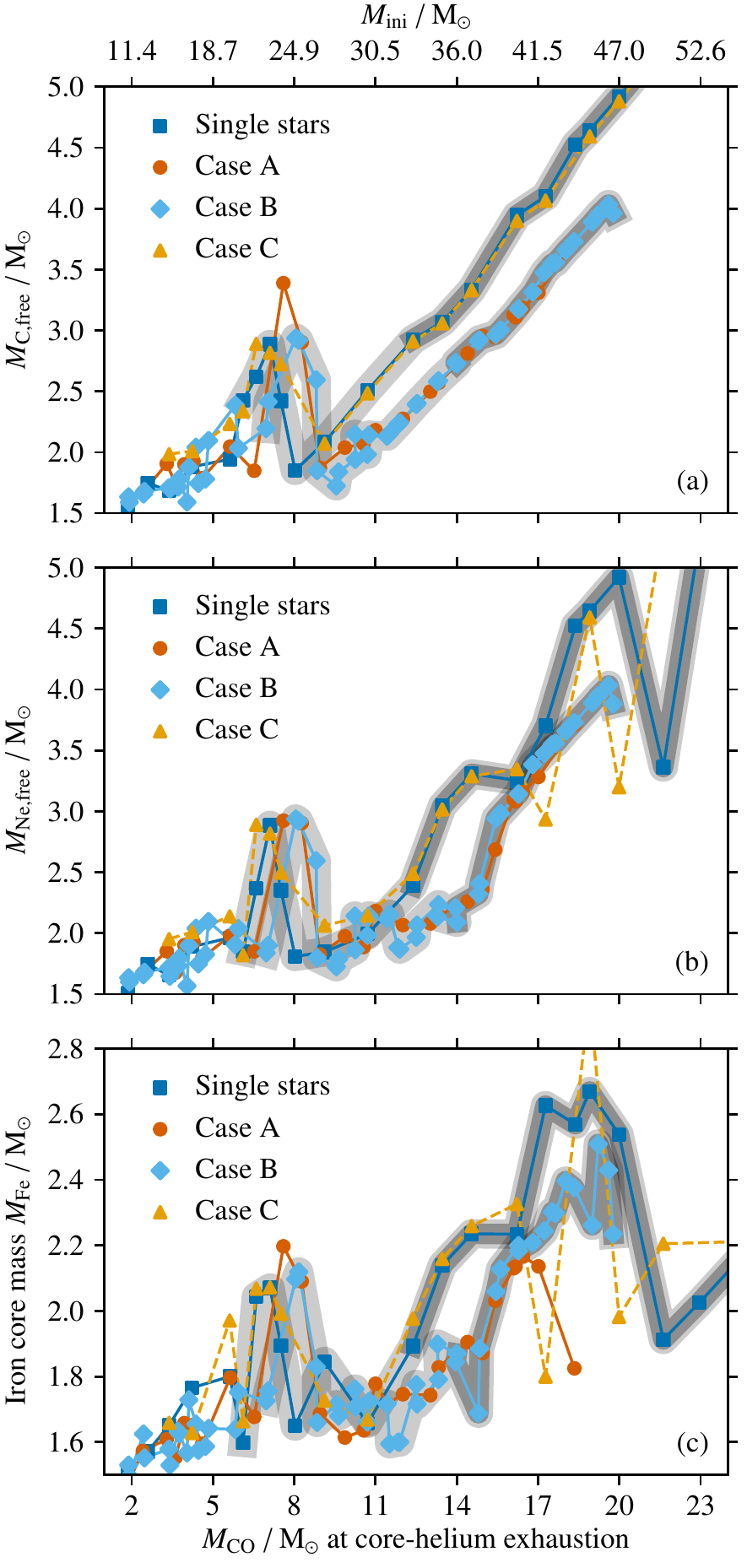} 
    \par\end{centering} \caption{Carbon-free (top panel a), neon-free (middle
    panel b) and iron core mass (bottom panel c) at core collapse as a function
    of CO core mass $M_\mathrm{CO}$. The carbon- and neon-free core masses are
    defined as those mass coordinates where the carbon and neon mass fraction
    falls below $10^{-5}$, \ie these core masses (mass coordinates) indicate the
    bottom of the carbon- and neon-burning layers at core collapse.}
    \label{fig:key-core-masses}
\end{figure}

The different starting points for the advanced-burning stages of single and
stripped stars shown in Fig.~\ref{fig:core-c-mass-fraction} persist up to the
pre-SN phase. We highlight these differences by considering the compactness
$\xi_{2.5}\equiv\xi_{M=2.5\,\msun}$ \citep{2011ApJ...730...70O},
\begin{equation}
    \xi_M = \frac{M/\msun}{R(M)/1000\,\km},
    \label{eq:xi}
\end{equation}
the dimensionless central specific entropy $s_\mathrm{c}$ and the iron core mass
$M_\mathrm{Fe}$ at core collapse
(Figs.~\ref{fig:compactness-and-central-entropy}
and~\ref{fig:key-core-masses}c). In Eq.~(\ref{eq:xi}), $M$ is the mass
coordinate at radius $R(M)$ at which the compactness is measured. All three
quantities are thought to be proxies for how likely it is that a star explodes
successfully in a supernova. Stars with a small compactness parameter, low
central entropy and small iron core mass tend to be more explodable and form a
NS.

The compactness $\xi_{2.5}$, entropy $s_\mathrm{c}$ and iron core mass
$M_\mathrm{Fe}$ are non-monotonic functions of $M_\mathrm{CO}$
(Figs.~\ref{fig:compactness-and-central-entropy}
and~\ref{fig:key-core-masses}c). We consider them as a function of
$M_\mathrm{CO}$ for easy comparison with the results in
Sect.~\ref{sec:single-vs-binary-star-evolution} and because $M_\mathrm{CO}$ is a
relatively easily accessible quantity that is also often used, \eg, in
population synthesis studies. They are essentially related to each other via the
pre-SN (iron) core mass. The compactness is directly connected to the
mass--radius relation of the ``core'', as is the central entropy. The pre-SN
iron cores have adiabatic profiles (\ie $s_\mathrm{c}=\mathrm{const.}$) such
that they can be described by $n=3/2$ polytropes. In such polytropes, the
polytropic constant $K$ scales as $K\propto M^3 R$ ($P=K\rho^{(n+1)/n}$) such
that the entropy is $s\propto \ln\, K \propto 3\ln\,M + \ln\,R$ with the mass
$M$ and the radius $R$ of the core. This implies that all three quantities
qualitatively trace the same shape as a function of $M_\mathrm{CO}$, as can
indeed be observed in Figs.~\ref{fig:compactness-and-central-entropy}
and~\ref{fig:key-core-masses}c. 

In our single stars and \casec mass-transfer systems, the compactness
$\xi_{2.5}$, entropy $s_\mathrm{c}$ and iron core mass $M_\mathrm{Fe}$ reach
local maxima at $M_\mathrm{CO}\,{\approx}\,7\,\msun$
(Figs.~\ref{fig:compactness-and-central-entropy}
and~\ref{fig:key-core-masses}c). All three quantities increase again for
$M_\mathrm{CO}\,{\gtrsim}\,11\,\msun$. The \casea and~B stripped stars follow a
similar, albeit shifted trend: the local maximum is now reached at
$M_\mathrm{CO}\,{\approx}\,8\,\msun$ and only increases again for
$M_\mathrm{CO}\,{\gtrsim}\,15\,\msun$. Both, the local maximum, referred to as
``compactness peak'' from hereon, and the increase in compactness for large
$M_\mathrm{CO}$ coincide with the transitions from convective to radiative
carbon and neon burning, respectively.

Regardless of the exact explosion mechanism, these results show that there is an
island of models around the compactness peaks of both the single and stripped
stars where a successful supernova explosion is less likely. Similarly for large
$M_\mathrm{CO}$ (${\gtrsim}\,11\,\msun$ for single and \casec stars, and
${\gtrsim}\,15\,\msun$ for \casea and~B stripped stars), it will become less
likely that stars can explode. In the range
$M_\mathrm{CO}=11\text{--}15\,\msun$, \casea and~B stripped stars at core
collapse have significantly smaller compactness parameters and iron core masses
than single stars, implying that they are more likely to explode. Consequently,
stripped stars in this $M_\mathrm{CO}$ range may lead to successful explosions
and NS formation whereas single stars might not explode and produce BHs.

To further understand this non-monotonic behaviour, we consider the iron core
masses as a function of $M_\mathrm{CO}$ (Fig.~\ref{fig:key-core-masses}). The
total mass of the pre-SN iron cores is set by how fast the previous nuclear
burning fronts move out in mass coordinate. Only those core regions that
completed carbon, neon, oxygen and silicon burning will constitute the iron
core. Especially the carbon- and neon-burning fronts limit the growth of the
iron cores as is illustrated in Fig.~\ref{fig:key-core-masses} by the
carbon-free and neon-free core masses at core collapse \citep[carbon- and
neon-free being defined as those regions where the respective mass fractions are
$<10^{-5}$; see also][]{2002ApJ...578..335F}. Changes in the speed with which
carbon and neon burning move out in mass coordinate lead to the
iron-core-mass/compactness/central-entropy peak and the increase in
$M_\mathrm{Fe}$, $\xi_{2.5}$ and $s_\mathrm{c}$ at $M_\mathrm{CO}\gtrsim
11\text{--}15\,\msun$.

Interestingly, \citet{2020arXiv200503055P} evolved bare CO cores of
$2.5\text{--}10\,\msun$ with initial carbon mass fractions of
$X_\mathrm{C}=0.05\text{--}0.50$ to core collapse and find a compactness
landscape as a function of $M_\mathrm{CO}$ and $X_\mathrm{C}$ that qualitatively
resembles our findings. The two branches of \casea and~B stripped stars, and
single and \casec stripped stars in Fig.~\ref{fig:core-c-mass-fraction} cross
the compactness landscape of \citet{2020arXiv200503055P} in different locations,
just as the single and helium star models in their figure 6.

In conclusion, the cores of \casea and~B stripped stars, and single and \casec
stripped stars evolve qualitatively similar up to core collapse. The SN outcome
of single and \casec stripped stars will of course differ despite the similar
core structures (\eg SN type, the hydrogen-rich envelope that may affect the SN
dynamics etc.; see next section).

\section{Core collapse and outcome}\label{sec:core-collapse-and-outcome}

\subsection{Compact-remnant masses}\label{sec:compact-remnant-masses}

\begin{figure*}
    \begin{centering}
    \includegraphics{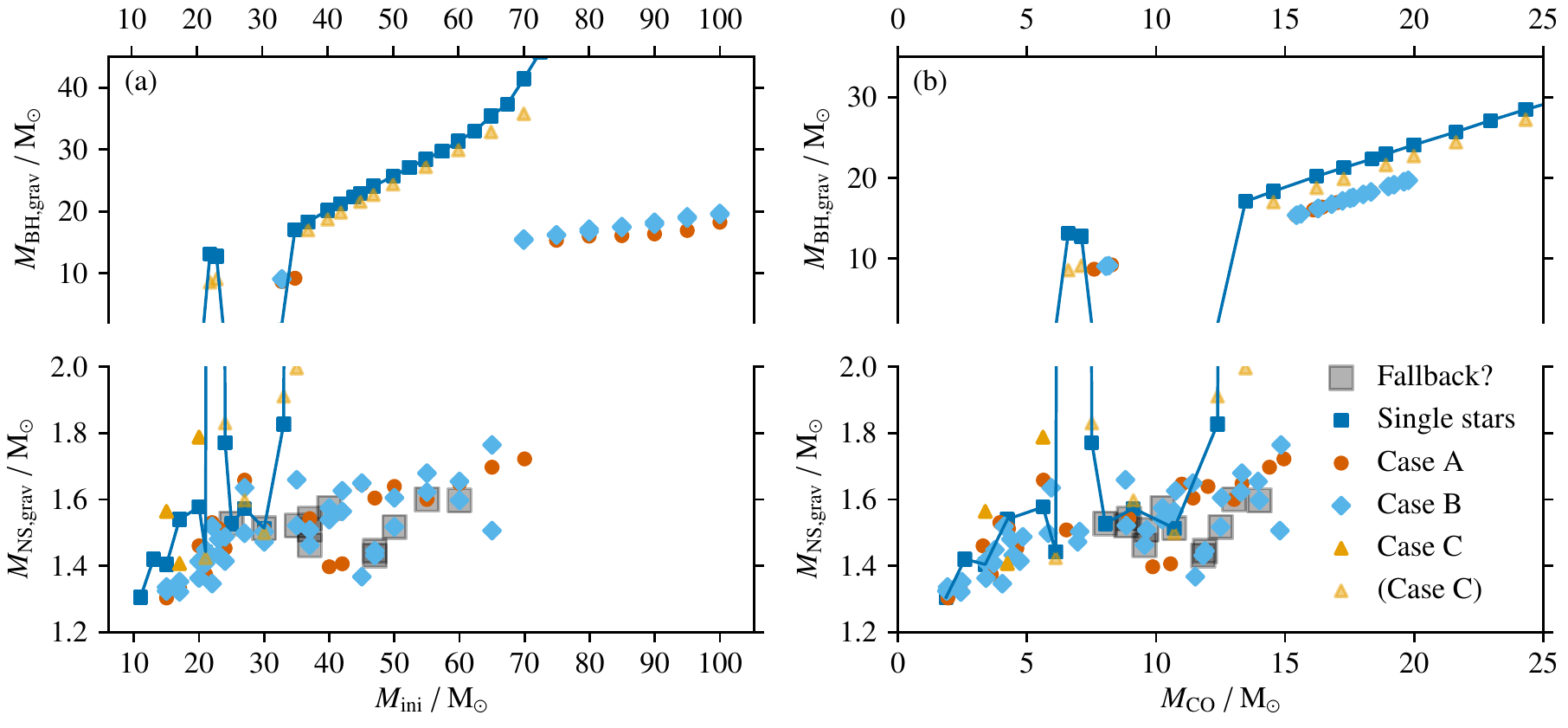} 
    \par\end{centering} \caption{Gravitational NS mass $M_\mathrm{NS,grav}$ and
    BH mass $M_\mathrm{BH,grav}$ as a function of the initial mass
    $M_\mathrm{ini}$ of stars (left panel a) and the CO core mass
    $M_\mathrm{CO}$ (right panel b). The large, gray boxes indicate models with
    potential (here unaccounted for) fallback that may result in weak or failed
    SNe. The light yellow triangles are for models that have reached their
    maximum radius before core helium burning and thus cannot undergo \casec
    mass transfer. They are shown here mainly for academic purpose.}
    \label{fig:compact-remnant-masses}
\end{figure*}

While the compactness parameter, the central entropy and the iron core mass may
be viewed as proxies of the explodability of stars, they cannot fully capture
the outcome of the complex core collapse of stars \citep[see
\eg][]{2015ApJ...801...90P, 2016ApJ...818..124E, 2016MNRAS.460..742M,
2020MNRAS.491.2715B}. Using the parametric SN code of
\citet{2016MNRAS.460..742M}, we predict the likely outcome of core collapse of
our models and show the resulting NS and BH masses as functions of initial mass
$M_\mathrm{ini}$ and CO core mass $M_\mathrm{CO}$ in
Fig.~\ref{fig:compact-remnant-masses}. We again use $M_\mathrm{CO}$ as measured
at the end of core helium burning to allow for direct comparisons with
Figs.~\ref{fig:core-c-mass-fraction}, \ref{fig:mini-mco},
\ref{fig:compactness-and-central-entropy} and~\ref{fig:key-core-masses}. The CO
core masses at core collapse are slightly larger than $M_\mathrm{CO}$ because of
helium shell burning (on average 0.3\%--1.4\% and at most 7\%).

In Fig.~\ref{fig:compact-remnant-masses}, we indicate those \casec systems that
within our assumptions would not occur in isolated binary-star evolution,
because the donor stars reach their maximum radius before core helium burning
(Sect.~\ref{sec:binary-star-models}). We also mark models that may be able to
launch an explosion by delayed neutrino heating, but where the energy input by
neutrinos is not sufficient to expel the entire stellar envelope according to
the semi-analytic SN model. A failed or weak supernova may result with
significant fallback of envelope material onto the formed compact remnant. Such
cases are found in both single and stripped stars beyond the respective
compactness peaks.

The predicted NS and BH masses from stripped stars are on average less massive
than those of single stars\footnote{This is less clear for NSs from \casec mass
transfer that rather scatter around the NS masses of single stars.} (when for
the moment not considering the complications due to fallback). Most importantly,
this is also true for single and stripped stars, even if they have the
\emph{same} CO core masses (Fig.~\ref{fig:compact-remnant-masses}).

\begin{figure*}
    \begin{centering}
    \includegraphics{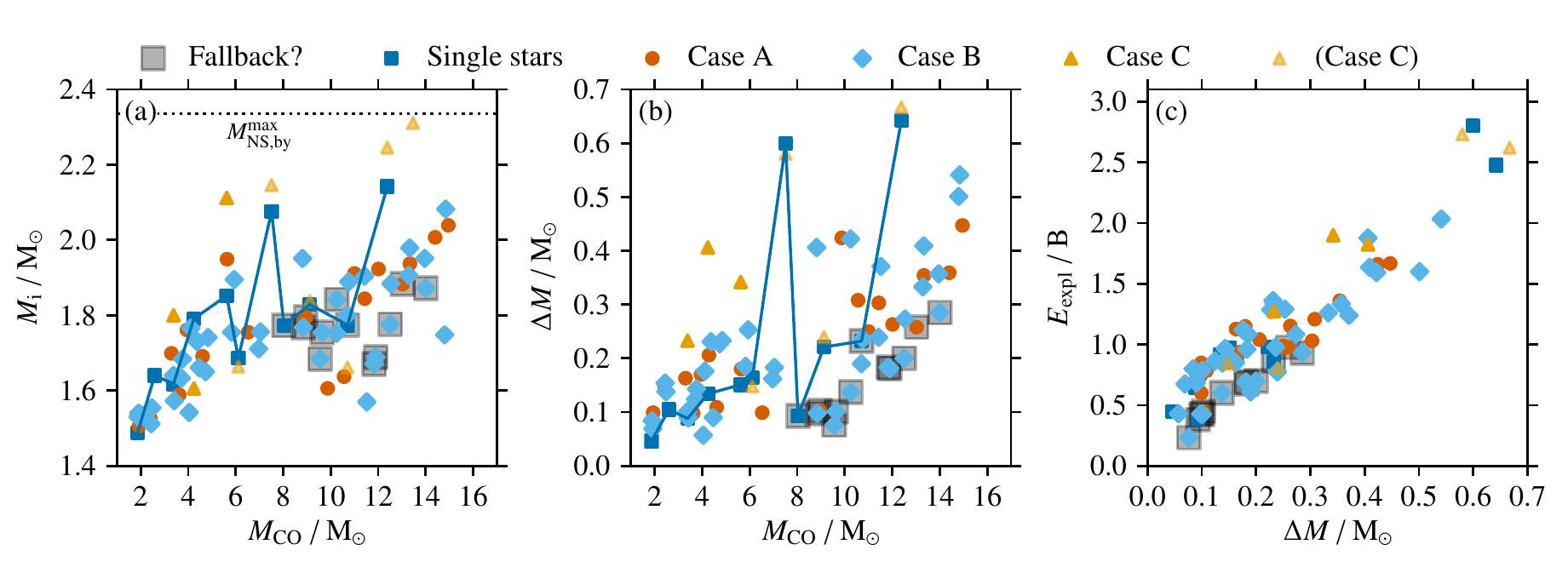} 
    \par\end{centering} \caption{Initial mass cut $M_\mathrm{i}$ of SN shock
    revival (panel a) and mass $\Delta M=M_\mathrm{f}-M_\mathrm{i}$ involved in
    accretion onto the proto-NS star (see Sect.~\ref{sec:parametric-sn-code}) as
    a function of $M_\mathrm{CO}$ (panel b), and the resulting SN explosion
    energy $E_\mathrm{expl}$ as a function of $\Delta M$ (panel c). The final NS
    mass correlates strongly with $M_\mathrm{i}$ (\cf
    Fig.~\ref{fig:compact-remnant-masses}b). The vertical dotted line in panel
    (a) is the assumed maximum (baryonic) NS mass.}
    \label{fig:Minit_Delta-M_Eexpl}
\end{figure*}

In NS formation, stripped stars encounter SN shock revival in our models at on
average smaller mass cuts $M_\mathrm{i}$ than single stars
(Fig.~\ref{fig:Minit_Delta-M_Eexpl}a), which can be partly understood from their
(on average) smaller iron cores (Fig.~\ref{fig:key-core-masses}c). The initial
mass cuts $M_\mathrm{i}$ are closely related to the final NS mass (\cf
Figs.~\ref{fig:compact-remnant-masses}b and~\ref{fig:Minit_Delta-M_Eexpl}a), and
this thus explains why stripped stars form on average lower NS masses. Also, the
amount of mass involved in the accretion onto the proto-NS, $\Delta M =
M_\mathrm{f} - M_\mathrm{i}$, ultimately determines the overall neutrino
luminosity and hence explosion energy (Fig.~\ref{fig:Minit_Delta-M_Eexpl}c).
This mass $\Delta M$ is again on average larger in stripped stars than in single
stars (Fig.~\ref{fig:Minit_Delta-M_Eexpl}b, excluding the fallback cases), so
stripped stars will give rise to on average more energetic SN explosions than
single stars. We come back to this and further consequences for the kicks and
nickel yields in Sects.~\ref{sec:explosion-energy-and-nickel-mass}
and~\ref{sec:kicks}. 

Note also that $M_\mathrm{i}$ and $\Delta M$ are closely related to the mass
coordinate of the base of the O shell, $M_4$, and the spatial mass derivative at
this point, $\mu_4$, respectively
(Appendix~\ref{sec:relation-SN-code-structure}), \ie to the two structural
parameters of pre-SN models that \citet{2016ApJ...818..124E} found to be good
predictors of the explodability of stars. In conclusion, stripped stars tend to
be ``easier'' to explode than single stars.

In BHs, the lower masses from stripped stars compared to single stars
(Fig.~\ref{fig:compact-remnant-masses}) are rather straightforward to understand
as the BH mass is given by the total final mass of stars. This mass is lower in
stripped stars that lost their hydrogen-rich envelopes than in single stars
(\casea stripped stars produce the lowest final masses, followed by \caseb and
\casec systems; for the possibility of a partial ejection of the hydrogen
envelope, see Sect.~\ref{sec:ns-bh-formation}). 

The compactness peaks described in Sect.~\ref{sec:pre-sn-structure}
(Fig.~\ref{fig:compactness-and-central-entropy}a) result in islands of BH
formation (Fig.~\ref{fig:compact-remnant-masses}). These islands are at
different masses for single stars and \casea\&~B stripped stars, analogously to
the shifted compactness peaks. While single stars of initially
${\approx}\,22\text{--}24\,\msun$ produce BHs of ${\approx}\,13\,\msun$ because
of the compactness peak, \casea and~B stripped stars do so for initial masses of
${\approx}\,32\text{--}35,\msun$ and give rise to BHs of ${\approx}\,9\,\msun$
(see also Table~\ref{tab:mini-nss}). So single stars in the compactness peak
produce BHs whereas stripped stars with the same CO core mass would
produce NSs and vice versa.

\begin{table}
    \caption{\label{tab:mini-nss}Initial mass ranges for NS formation in our single and stripped binary star models.}
    \centering
    \begin{tabular}{lccc}
    \toprule 
    Model & \multicolumn{3}{c}{Initial masses for NS formation} \\
    \midrule
    \midrule 
    Single stars & $M_\mathrm{ini}\lesssim 21.5\,\msun$ & and & ${\approx}\,23.5\text{--}34.0\,\msun$ \\
    \casea & $M_\mathrm{ini}\lesssim 31.5\,\msun$ & and & ${\approx}\,36.0\text{--}72.5\,\msun$ \\
    \caseb & $M_\mathrm{ini}\lesssim 31.5\,\msun$ & and & ${\approx}\,34.0\text{--}67.5\,\msun$ \\
    \casec & $M_\mathrm{ini}\lesssim 21.5\,\msun$ & and & ${\approx}\,23.5\text{--}36.0\,\msun$ \\
    \bottomrule
    \end{tabular}
\end{table}

Single stars of initially ${\gtrsim}\,35\,\msun$ produce BHs of
${\gtrsim}\,17\,\msun$ in our models (Fig.~\ref{fig:compact-remnant-masses} and
Table~\ref{tab:mini-nss}). Because envelope stripping tends to make it more
likely that a star can explode, our \casea and~B stripped stars may produce BHs
only for initial masses of ${\gtrsim}\,70\,\msun$
($M_\mathrm{CO}\gtrsim15\,\msun$). This is connected to the relatively moderate
compactness parameters of \casea\&~B stripped stars up to
$M_\mathrm{CO}\approx15\,\msun$ as shown in
Fig.~\ref{fig:compactness-and-central-entropy}a. However, our analysis indicates
that fallback may occur in some of the stripped stars in the CO core mass range
$9\text{--}15\,\msun$ such that these stars may then produce BHs. In fact, the
same is true for single stars beyond the compactness peak, and on average
20\%--40\% of our models beyond the compactness peak experience fallback and a
weak/failed SN. In conclusion, stripped stars produce NSs over a larger range of
initial and, even more importantly, over a larger range of CO-core masses than
single stars (Table~\ref{tab:mini-nss}).

There is a mass gap between NSs and BHs at about $2\text{--}9\,\msun$
(Fig.~\ref{fig:compact-remnant-masses}). This gap exists because the applied SN
mechanism only has two outcomes: successful explosion or formation of a BH. With
partial fallback of envelope material, the gap could be narrowed or even filled
completely. We will come back to this in the next section (see also discussion
in Sect.~\ref{sec:mass-gap}).

\subsection{NS--BH mass distribution}\label{sec:ns-bh-mass-distr}

\begin{figure*}
    \begin{centering}
    \includegraphics{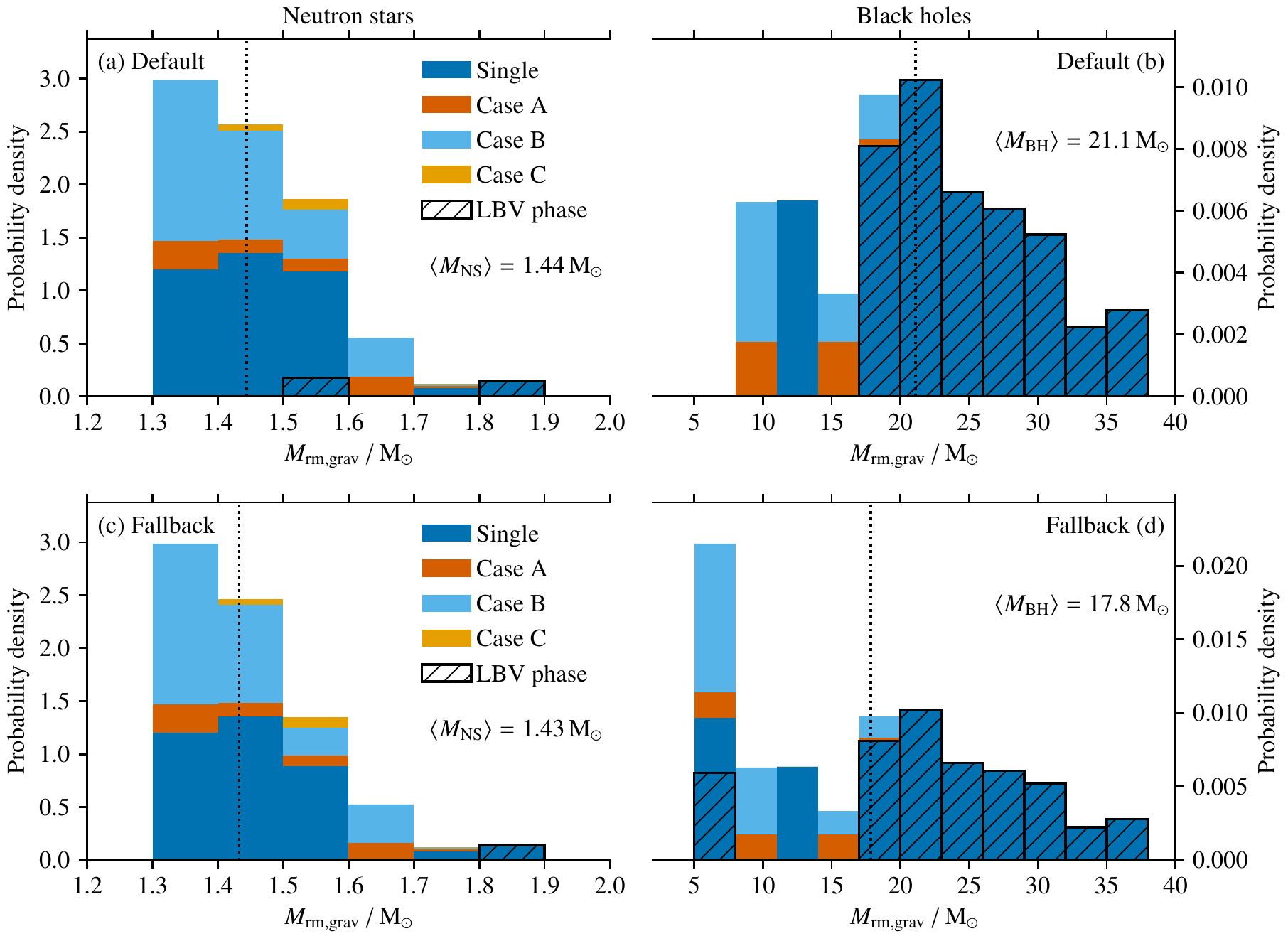} 
    \par\end{centering} \caption{Mass functions of NSs and BHs in our default
    model (top panel) and a model accounting for fallback (bottom panel). The
    vertical lines show the median masses of the NS and BH mass distributions,
    and the corresponding values are provided in the figure. Possible LBV phases
    are only indicated for the single stars.}
    \label{fig:comf-default-and-fallback-model}
\end{figure*}

We next construct NS and BH mass distributions for stellar populations born in a
single starburst at solar metallicity
(Fig.~\ref{fig:comf-default-and-fallback-model}). These distributions cannot be
directly compared to, \eg, compact-object mass distributions from
gravitational-wave observations, because the latter will be a mixture of stars
of different metallicities formed according to the cosmic star-formation
history. 

To set up the population, we assume that the initial masses of stars (in
binaries, the more massive primary star $M_1$) are distributed according to a
power-law, initial-mass function with high-mass slope $\gamma\,{=}\,-2.3$
\citep{1955ApJ...121..161S,2001MNRAS.322..231K,2003PASP..115..763C}. The
primordial binary fraction is set to 70\%, and uniform distributions of mass
ratio $q\,{=}\,M_2/M_1$ and logarithmic orbital separation $\log a$ are used
($M_2$ is the initial mass of the secondary star). Orbital separations are
limited to $\log a/\rsun\, {\leq}\, 3.3$ ($a\,{\lesssim}\,2000\,\rsun$). For
\casea, B and~C mass transfer, we assume that binaries merge and thus do not
result in a stripped star if $q\,{<}\,0.5$, $q\,{<}\,0.2$ and $q\,{<}\,0.2$,
respectively \citep[\cf][]{2015ApJ...805...20S}. We only consider genuine \casec
systems. Accounting for all binary interactions (mass losers, mass gainers,
mergers and CE systems), about 70\% of all stars in our population interact by
mass exchange in a binary and 30\% evolve as effective single stars, similarly
to what is expected in Galactic O stars \citep{2012Sci...337..444S}. 

In our sample, there are no stars that may explode in electron-capture
supernovae (ECSNe). Such SNe have been suggested to produce rather small NS
masses of ${\approx}\,1.25\,\msun$ \citep[\eg][]{2004ApJ...612.1044P,
2004ESASP.552..185V, 2010ApJ...719..722S} that are missing in our distributions.
Also, we only consider one binary-mass stripping episode, where additional such
episodes can occur and may lead to ultra-stripped stars that could then again
give rise to low-mass NSs \citep[maybe as low as $1.1\,\msun$,
see][]{2015MNRAS.451.2123T}. The lowest baryonic iron core mass in our sample is
$1.49\,\msun$ and the lowest gravitational NS mass is $1.32\,\msun$. 

In Sect.~\ref{sec:ns-bh-mass-distr-default}, we first discuss the compact-object
mass distributions of our default stellar-remnant masses
(Fig.~\ref{fig:compact-remnant-masses}) and, in
Sect.~\ref{sec:ns-bh-mass-distr-fallback}, we then also consider partial fallback.

\subsubsection{Default model}\label{sec:ns-bh-mass-distr-default}

First, we describe the results of the genuine single stars
(Fig.~\ref{fig:comf-default-and-fallback-model} top panel). Their NS masses
follow a unimodal distribution with a tail up to $1.9\,\msun$ (the largest NS
mass in the single-star sample is $1.87\,\msun$). The maximum allowed NS mass in
this study is $2.0\,\msun$, but it is not reached by stars in our grid of
models. A finer grid is probably required to sample the NS mass distribution up
to the highest masses. In any case, some NSs are born massive in our models, in
agreement with pulsar observations  \citep[\eg][]{2010Natur.467.1081D,
2011ApJ...732...70L, 2011MNRAS.416.2130T}. High individual NS masses are also
inferred in the compact object merger GW190425 with a total mass of
${\approx}\,3.4\,\msun$ \citep{2020ApJ...892L...3A}. 

The NS-mass distribution has a gap or dearth at $1.6\text{--}1.7\,\msun$
(Fig.~\ref{fig:comf-default-and-fallback-model}a), which is caused by stars in
the compactness peak for which we predict BH formation (see the ``BH island'' at
$M_\mathrm{CO}\,{\approx}\,7\,\msun$ in Fig.~\ref{fig:compact-remnant-masses}
and the corresponding compactness peak in
Fig.~\ref{fig:compactness-and-central-entropy}a). Whether the NS mass
distribution shows a gap, dearth or other feature because of the compactness
peak also depends on the NS masses of other stars (see
Appendix~\ref{sec:comf-toy-model}): for example, there is no gap or dearth if
stars beyond the compactness peak give rise to NS masses that are lower than
those of stars before the compactness peak.

The BH-mass distribution of single stars is bimodal
(Fig.~\ref{fig:comf-default-and-fallback-model}b). The first peak at
${\approx}\,12.5\,\msun$ is from stars in the compactness peak that failed to
explode. The second, major contribution to the BH-mass distribution starts at a
BH mass of ${\approx}\,17\,\msun$. This mass is set by the final mass of the
lowest-mass single star beyond the compactness peak that fails to explode. This
mass is mainly set by the total (wind) mass loss of a star, which implicitly
depends on the exact evolutionary history of the star and parameters such as
convective-boundary mixing (\eg convective overshooting). The most massive BH in
our sample (${\approx}\,50\,\msun$; not visible in
Figs.~\ref{fig:compact-remnant-masses}
and~\ref{fig:comf-default-and-fallback-model}) is essentially given by the final
mass of the most massive single star that we evolved up to core collapse (here
$75\,\msun$). Technically, this could be larger than what we have in our current
sample, but our models do not include mass loss during LBV phases or other forms
of pulsational and eruptive mass loss from massive stars \citep[see
\eg][]{2014ARA&A..52..487S}, which likely limits BH masses from single stars. 

We indicate stars that potentially experience LBV-like mass loss by the black
hatching in Fig.~\ref{fig:comf-default-and-fallback-model} and their compact
remnant masses should thus rather be considered as upper limits. Stars are
indicated to undergo enhanced mass loss if they cross the hot side of the
S~Doradus instability strip \citep[\eg][]{2004ApJ...615..475S} or a luminosity
of $\log L/\lsun=5.5$ \citep{2018MNRAS.478.3138D} for effective temperatures
$T_\mathrm{eff}<14,350\,\mathrm{K}$, \eg as red supergiants.

The NS and BH mass distributions of stripped stars show qualitatively similar
features as those of genuine single stars. However, there are important
quantitative differences. First, the compactness peak is at larger CO core
masses and does not result in a very pronounced dearth in the NS mass
distribution (Fig.~\ref{fig:comf-default-and-fallback-model}a). The gap in the
NS mass distribution produced by the compactness peak in single stars is now
filled with NS from stripped binary stars. 

Secondly, the NS masses of stripped stars are on average smaller than those of
single stars (see also Fig.~\ref{fig:compact-remnant-masses}b). The average NS
mass of our stripped binary models is ${\approx}\,1.42\,\msun$, while it is
${\approx}\,1.46\,\msun$ in the single star models. This systematic difference
is further exemplified by the NS mass distribution of stripped binary stars that
peaks at ${\approx}\,1.35\,\msun$ while that of single stars peaks at
${\approx}\,1.45\,\msun$. From our models, it also seems that the NS mass
distribution of stripped stars is effectively truncated at about $1.6\,\msun$,
whereas that of single stars extends up to $1.9\,\msun$. The NSs born with
masses ${\gtrsim}\,1.6\text{--}1.7\,\msun$ in the single and stripped stars are
from stars close to the transitions from NS to BH formation. Because these
transitions are at lower initial masses for single stars, there are relatively
more such massive NSs because of the IMF and a more pronounced tail is formed
towards large NS masses.

Thirdly, the maximum BH mass of stripped stars in our sample is
${\approx}\,20\,\msun$ compared to ${\approx}\,50\,\msun$ in single stars
(Fig.~\ref{fig:comf-default-and-fallback-model}b). This maximum mass could be
slightly larger when considering stripped binary models from initially
${>}\,100\,\msun$ stars. Also, the contributions of BHs from stars in and beyond
the compactness peak, \ie BH masses of ${\approx}\,9\,\msun$ and
${\gtrsim}\,15\,\msun$, respectively, are lower in stripped stars in comparison
to single stars with ${\approx}\,12.5\,\msun$ and ${\gtrsim}\,17\,\msun$,
respectively. The main difference in the masses is of course due to the stripped
envelope in the binary models.

There are no BHs in the mass range ${\approx}\,2.0\text{--}8.7\,\msun$ in our
default model of stripped binary stars
(Fig.~\ref{fig:comf-default-and-fallback-model} top panel). As mentioned above,
this is because in our SN model, we assume either a complete explosion without
any fallback or no explosion with total fallback of the whole final stellar
mass. We relax this assumption in the next section.

\begin{figure*}
    \begin{centering}
    \includegraphics{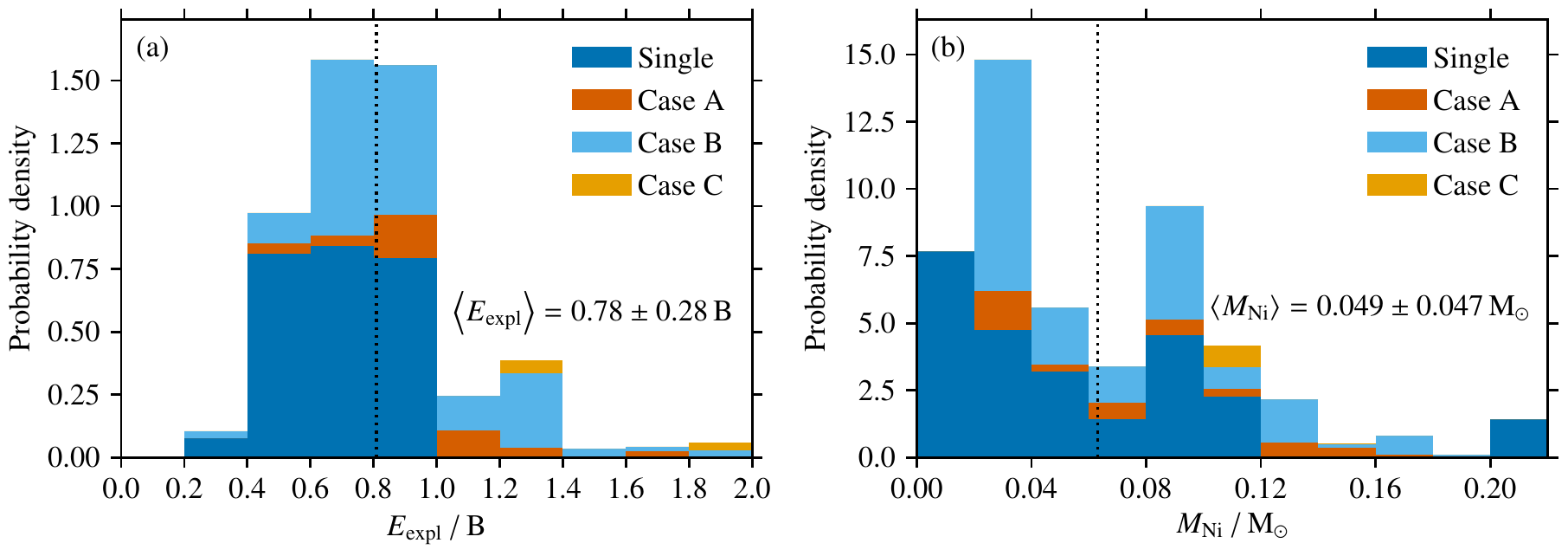} 
    \par\end{centering} \caption{Distribution of (a) explosion energies and (b)
    nickel masses of single and stripped binary stars.}
    \label{fig:explosion-energy-and-nickel-mass}
\end{figure*}

\subsubsection{Fallback model}\label{sec:ns-bh-mass-distr-fallback}

Within the applied SN model, we encounter cases where an explosion is triggered,
but it is then not energetic enough to explode the whole star (the fallback
cases in Fig.~\ref{fig:compact-remnant-masses}). We now assume that 50\% of the
presumed ejecta mass falls back and adds to the compact remnant mass. Instead of
forming NSs, BHs are produced. 

The fallback narrows down the mass gap between NSs and BHs from
${\approx}\,2.0\text{--}8.7\,\msun$ to ${\approx}\,2.0\text{--}5.4\,\msun$
(Fig.~\ref{fig:comf-default-and-fallback-model} bottom panel). A smaller
fallback fraction leads to lower compact-object masses (\ie BHs) and would thus
further narrow down the gap. Vice versa, a larger fallback fraction would widen
it. In principle, the gap could disappear altogether, in particular if various
fallback fractions are realized in nature.

The main characteristics of the NS and BH mass distributions (\cf
Sect.~\ref{sec:ns-bh-mass-distr-default}) remain intact otherwise, \eg stars in
the compactness peak still make a distinct contribution to the BH distribution
and may lead to a dearth in the NS distribution. Because the fallback BHs stem
from on average lower initial masses than other BHs, they add a significant
contribution to the BH mass distribution at ${\approx}\,5.0\text{--}8.0\,\msun$,
\ie they are the least massive BHs in our models. This contribution is larger
than that from BHs formed from stars in the compactness peak. While not
immediately visible in Fig.~\ref{fig:comf-default-and-fallback-model}d, the
lowest-mass fallback BHs are still from stripped stars, but this may no longer
be true when considering that different stars may experience varying levels of
fallback.

\subsection{Explosion energy and nickel mass}\label{sec:explosion-energy-and-nickel-mass}

As described in Sect.~\ref{sec:parametric-sn-code}, we re-calibrated the
parametric SN code of \citet{2016MNRAS.460..742M} such that the average
explosion energy of SN~IIP is $0.69\pm0.17\,\mathrm{B}$ (stripped binary models
do not contribute to this calibration). Using the same population model as in
Sect.~\ref{sec:ns-bh-mass-distr}, we show the distribution of explosion energies
of our single and stripped binary stars in
Fig.~\ref{fig:explosion-energy-and-nickel-mass}a. The average explosion energy
of SN~Ib/c is $0.88\pm0.31\,\mathrm{B}$ (almost exclusively stripped binary
stars) and is systematically larger by 28\% than that of SN~IIP.

\begin{figure*}
    \begin{centering}
    \includegraphics{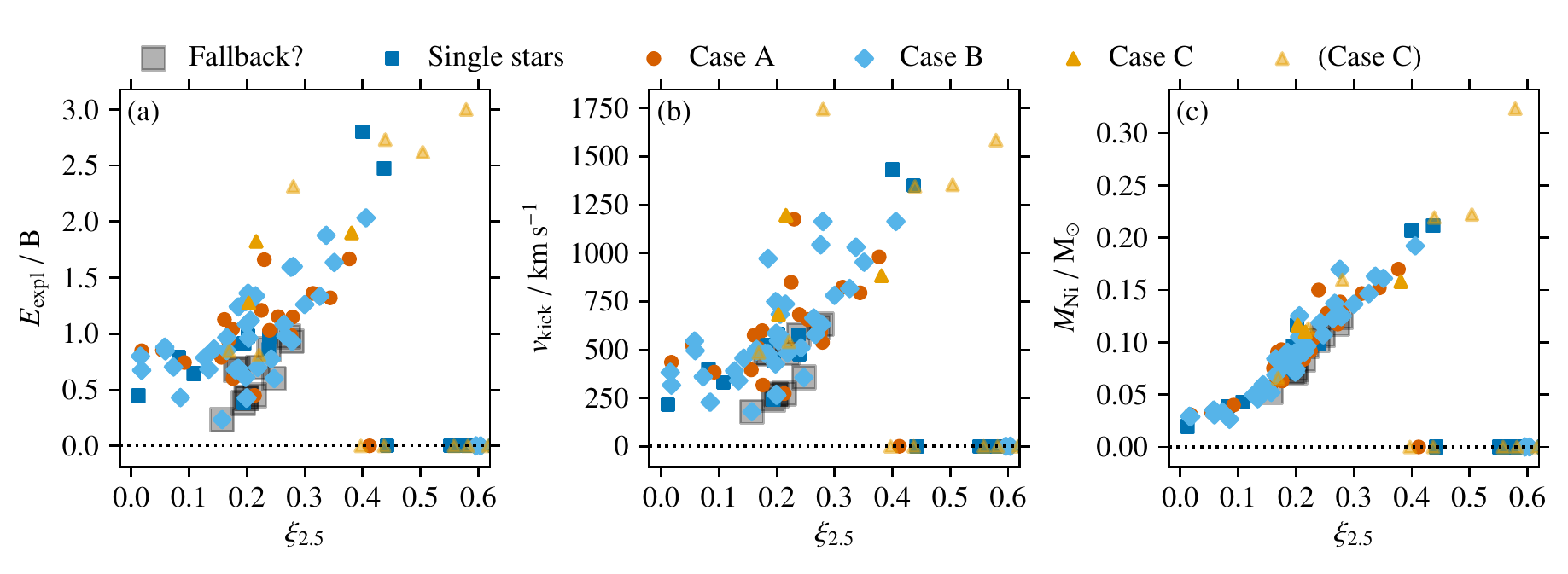} 
    \par\end{centering} \caption{Explosion energy (a), kick velocity (b) and
    nickel mass (c) as a function of compactness.}
    \label{fig:Eexpl-vkick-MNi-vs-compactness}
\end{figure*}

The more energetic explosions of stripped stars in our models are a consequence
of the larger mass $\Delta M$ that is accreted onto the NS and powers the
neutrino luminosity (Fig.~\ref{fig:Minit_Delta-M_Eexpl}c) as discussed in
Sect.~\ref{sec:compact-remnant-masses}. Another often used proxy for the
explosion energy is the compactness parameter $\xi_{2.5}$
\citep[Fig.~\ref{fig:Eexpl-vkick-MNi-vs-compactness}a; see
also][]{2015PASJ...67..107N, 2016MNRAS.460..742M}, but it appears to be of
limited use in our models. We find that models with $\xi_{2.5}\lesssim0.15$ have
a similar explosion energy of about $0.75\,\mathrm{B}$ and that there is an
approximately linear trend of explosion energy with compactness for
$\xi_{2.5}\gtrsim0.15$. However, the scatter around this general trend is large
and precludes the use of the compactness as a predictive quantity for explosion
energy. For $\xi_{2.5}\lesssim0.40$, we exclusively find SN explosions, while
both successful and failed SNe are possible for more compact stars.

More energetic explosions also result in larger nickel yields
(Fig.~\ref{fig:explosion-energy-and-nickel-mass}b) and SN kicks
(Sect.~\ref{sec:kicks}) as is also evident from the general correlations of
compactness with nickel yield and SN kick velocity
(Fig.~\ref{fig:Eexpl-vkick-MNi-vs-compactness}). The correlation of nickel yield
and compactness shows the least scatter. This suggests that the nickel yields
from observations could shed light on the compactness of an exploding star,
although further investigation with more detailed SN models is warranted.

Because of the more energetic explosions of stripped stars, we find that SN~Ib/c
produce on average $0.059\,\msun$ nickel, while SN IIP produce on average
$0.039\,\msun$. Taken together, we find an average nickel yield of
$0.049\,\msun$ (Fig.~\ref{fig:explosion-energy-and-nickel-mass}b). Higher nickel
masses in more energetic explosions can be understood as follows. The higher
explosion energies lead to higher shock temperatures over a larger range in the
stellar interior, which enables more nuclear burning that produces nickel.

\subsection{NS and BH kicks}\label{sec:kicks}

\begin{figure*}
    \begin{centering}
    \includegraphics{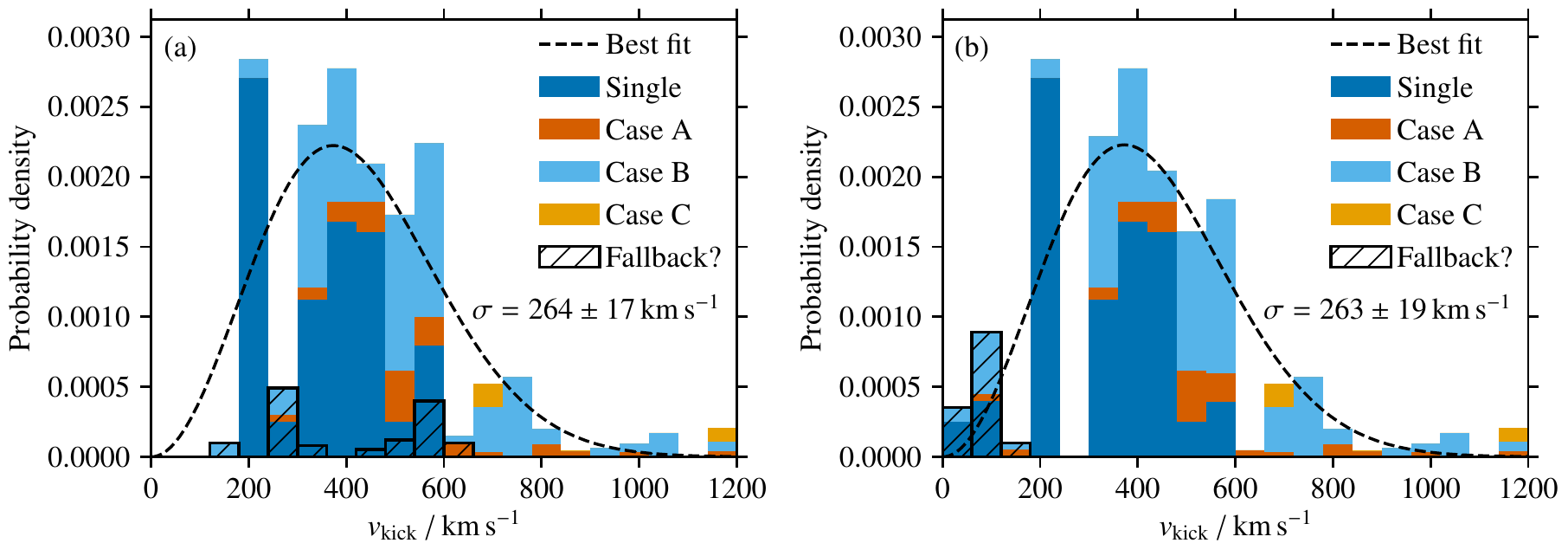} 
    \par\end{centering} \caption{Supernova-kick distribution of all stars
    without (left panel a) and with (right panel b) fallback. Maxwellian
    distributions are fitted to the data, and their best fit $\sigma$-values are
    provided.}
    \label{fig:sn-kick-distributions}
\end{figure*}

As is done in Sect.~\ref{sec:ns-bh-mass-distr} and
~\ref{sec:explosion-energy-and-nickel-mass}, we compute the distribution of the
mean SN kick velocities from our population of single and stripped binary
models. We consider the default and fallback cases separately
(Fig.~\ref{fig:sn-kick-distributions}). In the fallback case, we assume that the
NS of mass $M_\mathrm{NS}$ has received a kick of $v_\mathrm{kick}$ before it
accretes fallback material. The new kick of the compact object (most likely a
BH) then follows from linear momentum conservation,
\begin{equation}
    v_\mathrm{kick,new} = \frac{M_\mathrm{NS}}{M_\mathrm{NS} + M_\mathrm{fallback}} v_\mathrm{kick},
    \label{eq:vkick-new}
\end{equation}
where $M_\mathrm{fallback}$ is the amount of fallback mass. As detailed in
Sect.~\ref{sec:ns-bh-mass-distr-fallback}, we here assume that the fallback
material is 50\% of the original ejecta mass. 

Note that all our kick velocities are mean values and do not take the stochastic
nature of SN kicks into account (see Sect.~\ref{sec:parametric-sn-code}). Also,
the kick velocities in Fig.~\ref{fig:sn-kick-distributions}a are calibrated to
match the observed distribution of \citet{2005MNRAS.360..974H}, \ie a Maxwellian
distribution with $\sigma=265\,\kms$. In our models, the kick velocity scales
directly with explosion energy, the matter involved in the accretion onto the
proto-NS and the mass of the NS itself, (Eq.~\ref{eq:vkick}),
$v_\mathrm{kick}\propto\sqrt{\Delta M E_\mathrm{expl}}/M_\mathrm{NS,grav}$. Our
previous findings of on average higher $E_\mathrm{expl}$, larger $\Delta M$ and
smaller $M_\mathrm{NS,grav}$ in stripped stars compared to single stars
immediately imply that also the kicks within our SN model are on average larger
in stripped stars, as shown in Fig.~\ref{fig:sn-kick-distributions}.

Quantitatively, mean kick velocities are in the range $180\text{--}1500\,\kms$
($50\text{--}1500\,\kms$) without (with) fallback. The best-fitting Maxwellian
distributions are for $\sigma=222\pm23\,\kms$ and $\sigma=315\pm24\,\kms$ in
single and stripped stars, respectively\footnote{The best-fitting Maxwellian
distributions are for $\sigma=218\pm24\,\kms$ and $\sigma=321\pm27\,\kms$ in the
50\% fallback model.}. 

When plotting the kick velocity against the compactness of pre-SN stars, we see
the same qualitative trend as found in the explosion energy
(Fig.~\ref{fig:Eexpl-vkick-MNi-vs-compactness}b): successful explosions from
stars with a higher compactness lead to larger explosion energies and hence kick
velocities. Analogously to the explosion energies, the largest kicks are found
in stripped binary stars and almost all kicks larger than $600\,\kms$ are from
stripped binary models.

Models likely experiencing fallback are marked in
Fig.~\ref{fig:sn-kick-distributions} and all of the BHs formed by fallback in
our models receive (mean) kick velocities in the range
${\approx}\,50\text{--}150\,\kms$, \ie significantly slower than most of the
kick velocities of NSs. As shown in Sect.~\ref{sec:ns-bh-mass-distr-fallback},
the fallback BHs populate the mass gap between NSs and non-fallback BHs. In our
models, fallback BHs would always receive a kick, which might not be the case
for more massive BHs that may form by direct collapse. If the fallback fraction
is less than 50\% of the ejecta mass, the final mass of the compact remnant will
decrease and the kick velocity will increase (and vice versa). For a
distribution of fallback fractions as might be expected in reality, this implies
that the average kick velocity would decrease with increasing BH mass.

\section{Compact-object populations and merger statistics}\label{sec:pop-syn}

Using a toy population-synthesis model, we now study how binary-mass stripping
may affect compact-object populations and particularly merger rates. The toy
model is meant to be simple and to offer guidance on the expected
order-of-magnitude changes in NS-NS, BH-NS and BH-BH merger rates because of
envelope stripping in binary stars. It will not be able to properly catch all of
the complex intricacies of a full population-synthesis computation, which is
left for future work. For example, we only consider a single starburst
population of stars of solar metallicity without accounting for the cosmic
star-formation history. Metallicity-dependent stellar winds directly affect the
final masses of stars and possibly also the SN outcome, and the masses of
compact objects are important for the detection rates of merger events. We
describe the population set-up and assumptions in
Sect.~\ref{sec:toy-pop-syn-model}, compare a few population predictions to
Galactic compact objects in Sect.~\ref{sec:co-comparison} and present our
findings on compact-object merger rates in Sect.~\ref{sec:merger-rates} and
chirp-mass distributions in Sect.~\ref{sec:chirp-mass-distribution}.

\subsection{Toy population model}\label{sec:toy-pop-syn-model}

A typical binary system leading to a double compact-object merger within a
Hubble time undergoes the following key steps that we try to capture in our
population model. The first mass transfer episode (\eg \caseb) from the primary
to the secondary star is stable. The primary star is stripped off its envelope,
directly modifying its core evolution and hence compact-object remnant as shown
in this work. The secondary star accretes mass and rejuvenates. If the binary
survives the first SN kick, it may undergo a second mass-transfer phase that may
now be unstable and lead to a common-envelope phase. During this phase (and
possibly thereafter in another mass-transfer episode), the initial
secondary star is also stripped off its envelope with consequences for its final
fate and compact remnant. During the common-envelope phase, the orbit greatly
shrinks and the binary star has a higher chance to survive the kick from the
second SN. Ultimately, the two compact objects merge in a Hubble time thanks to
gravitational-wave emission. Such a channel makes up about 70\% of NS-NS mergers
in the models of \citet{2018MNRAS.481.4009V}.

As in Sect.~\ref{sec:ns-bh-mass-distr}, primary-star masses $M_1$ are sampled
from a power-law IMF with $\gamma=-2.3$ and secondary-star masses from a uniform
mass-ratio distribution (the minimum companion mass is $1\,\msun$). Initial
masses are limited to $M_\mathrm{max}=70\,\msun$ to account for the fact that
wind mass loss and enhanced mass loss from LBV-like eruptions and pulsations
widen orbits such that binary systems do not experience mass exchange anymore
\citep[\eg][]{1991A&A...252..159V,2015ApJ...805...20S}. For the same reason, we
limit BH masses for our solar metallicity models to at most $25\,\msun$. Orbital
separations are limited to $\log a_\mathrm{max}=3.3$ as before and are sampled
from a uniform distribution in $\log a$. We assume that binary stars undergoing
\casea mass transfer do not contribute to compact-object mergers and thus only
consider \caseb and~C systems. This is because the orbital periods after \casea
mass transfer are rather short and the subsequent common-envelope phase likely
leads to a merger \citep[\eg][]{1995ApJ...445..367T}. After the first
mass-transfer episode from the primary to the secondary star, we assume that the
secondary accretes 40\% of the mass of the primary star and that it rejuvenates
in the sense that its evolution after mass accretion can be well described by
its new total mass. We do not follow the exact orbital evolution of binary
stars, \ie also not through common-envelope phases.

The initial mass ranges for white-dwarf (WD) and NS formation are different in
single and binary-stripped stars, and are closely connected to the occurrence of
ECSNe\footnote{To be more precise, we not necessarily only have ECSNe in mind,
but also collapsing stars with very small iron cores that are not expected to
produce large SN kicks \citep{2004ApJ...612.1044P}. Such stars are commonly
produced in \casebb mass transfer and are referred to as ultra-stripped stars
\citep{2015MNRAS.451.2123T}. With this definition in mind, the initial mass
range over which low SN kicks are found could even be larger than what we assume
here.}. In single (stripped) stars, we assume WD formation for
$M_\mathrm{ini}<9.5\,\msun$ ($M_\mathrm{ini}<10.5\,\msun$). Also the initial
mass range of ECSNe is expected to be different in stripped binary stars
compared to single stars \citep[\eg][]{2004ApJ...612.1044P, 2017ApJ...846..170T,
2017ApJ...850..197P}. We assume that single stars give rise to ECSNe for initial
masses in the range $9.5\text{--}10.0\,\msun$, while this mass range is
$10.5\text{--}12.0\,\msun$ in stripped binary stars. The exact mass ranges are
currently uncertain, which we find to be important in particular for the NS-NS
merger rate. Here, we want to capture the general expectation that stripped
binary stars likely produce ECSNe over a larger initial mass range than single
stars and that higher initial masses are required because of envelope stripping.

Supernova kicks are a key (yet considerably uncertain) ingredient in predicting
compact-object-merger rates \citep[\eg][]{2018MNRAS.480.2011G}. Within our toy
model, we reduce the number of binaries that can lead to compact-object mergers
for each kick and apply different reduction factors depending on whether NSs or
BHs are formed. For NSs formed in ECSNe, we assume that binaries are not
disrupted. For NSs formed in CCSNe, the first and second SNe are assumed to
break up 90\% and 20\% of binaries, respectively. The latter two fractions are
average values reported by \citet{2018MNRAS.481.4009V} in their more detailed
population synthesis work. Also \citet{2019A&A...624A..66R} report a similar
break-up fraction of $86^{+11}_{-9}\%$ for the first SN in a binary system. For
BHs, we assume that 20\% and none break up at the first and second SN,
respectively. If the BH formed by fallback, we increase the break-up fraction to
40\% for the first SN.

In our models we find that 20\%--40\% of stars beyond the compactness peak may
experience fallback and thereby BH formation. We therefore assume that one third
of models beyond the compactness peak that may form NSs will experience fallback
of 50\% of the ejecta mass (as assumed in
Sect.~\ref{sec:ns-bh-mass-distr-fallback}). 

Initial masses are related to CO core masses and hence NS\&BH masses through
fitting formulae to our single-star, and \casea, B and~C stripped binary-star
models (Fig.~\ref{fig:mini-mco} and Appendix~\ref{sec:fit-functions}). NSs from
ECSNe are all assumed to have a mass of $1.25\,\msun$
\citep{2010ApJ...719..722S}. Below, we also consider the rapid and delayed
supernova model of \citet{2012ApJ...749...91F} for comparison, because it is
regularly employed in state-of-the-art population synthesis computations of
gravitational-wave sources (\eg in StarTrack, \citealt{2020A&A...636A.104B},
Compas, \citealt{2017NatCo...814906S}, and MOBSE,
\citealt{2020ApJ...891..141G}), but there are also other population synthesis
models that use different prescriptions \citep[\eg
COMBINE,][]{2018MNRAS.481.1908K}.

In our population model, we have made an implicit assumption, namely that
compact-object mergers form similarly from binaries with different primary star
masses (\ie the progenitors of NS-NS and BH-BH mergers follow the same
evolutionary paths). Qualitatively, this may not be such a bad assumption, but
it must not necessarily hold quantitatively. For example, the fraction of binary
systems experiencing an unstable first mass-transfer episode likely decreases
with primary mass, implying that the fraction of binary systems undergoing CE
evolution is lower in more massive primary stars \citep[see
\eg][]{2015ApJ...805...20S}. Our toy model captures part of this, because the
available parameter space for \caseb and~C mass transfer is smaller in more
massive primary stars. Furthermore, we do not include the close binary channel
invoking chemically homogenous evolution for the formation of massive BH-BH
mergers \citep{2016MNRAS.458.2634M, 2016A&A...588A..50M} in our toy model, and
also not dynamical formation channels \citep[\eg][]{2015PhRvL.115e1101R,
2016MNRAS.459.3432M, 2017MNRAS.467..524B}.

In the following, we make a differential analysis and only consider relative
quantities. In particular, we consider the following models that mainly differ
in their mapping from initial to compact-remnant masses:

\begin{itemize}
    \item \emph{Single}: The final fate (\ie ECSN, CCSN and NS or BH formation)
    and the compact remnant mass of the primary and secondary star in binaries
    are according to our single star models (see Figs.~\ref{fig:mini-mco}
    and~\ref{fig:com-toy-models}a). This is known to be a particularly bad
    assumption and is considered here only for reference. 
    \item \emph{Default}: This is our default model. We map the initial masses
    of stars to CO core masses and then to compact object masses using our
    \caseb models (see Figs.~\ref{fig:mini-mco} and~\ref{fig:com-toy-models}c).
    \item \emph{CPS}: This and the following two models are our attempt to mimic
    what is done in current population-synthesis (CPS) models. As in our default
    model, we map the initial masses of the primary and secondary star to CO
    core masses using our \caseb stripped binary star models, but the mapping
    from CO core mass to final fate (\eg NS or BH formation) is according to our
    single star models. If a NS is formed, we use the NS mass predicted by the
    single-star mapping, and BH masses are equal to the final mass of the \caseb
    stripped model.
    \item \emph{F12 rapid}: As is in the CPS model, but the mapping from pre-SN
    CO core mass to compact remnant mass is according to the rapid explosion
    model described in \citet{2012ApJ...749...91F}.
    \item \emph{F12 delayed}: As is in F12 rapid, but now using the delayed
    explosion model of \citet{2012ApJ...749...91F}.
\end{itemize}

\subsection{Comparison to Galactic compact objects}\label{sec:co-comparison}

In Table~\ref{tab:pop-syn}, we summarise a few key quantities of the models
described in Sect.~\ref{sec:toy-pop-syn-model}. The NS to BH ratio is computed
for stars initially up to $100\,\msun$. The NS to BH ratio in our
stripped-binary models is significantly higher than in the single star models
and the two F12 models. This is because the envelope stripping in our binary
models greatly extends the initial mass range over which NS and not BH formation
is found (Table~\ref{tab:mini-nss}). While the differences in the NS to BH
ratios are large, they appear less drastic when considering the fraction of NSs
among all NSs and BHs ($N_\mathrm{NS}$ and $N_\mathrm{BH}$ being the number of
NSs and BHs, respectively), 
\begin{equation}
    \frac{N_\mathrm{NS}}{N_\mathrm{NS}+N_\mathrm{BH}} = \frac{f_\mathrm{NS/BH}}{f_\mathrm{NS/BH}+1} \approx 75\% \text{--} 94\%
    \label{eq:NS-fraction}
\end{equation}
for $f_\mathrm{NS/BH}=3\text{--}15$ (Table~\ref{tab:pop-syn}). In particular,
the NS fraction is only three per-cent points higher in our default stripped
binary model (94\%) than in the CPS model (91\%), despite the quite large
difference in the NS to BH ratio and the fact that about 1/3 of the BHs in the
CPS model are NSs in the default model.

\setlength{\tabcolsep}{4pt}
\begin{table}
    \caption{\label{tab:pop-syn}Fractions of NSs to BHs ($f_\mathrm{NS/BH}$), single NSs ($f_\mathrm{NS}^\mathrm{s}$), single BHs ($f_\mathrm{BH}^\mathrm{s}$), single radio pulsars ($f_\mathrm{rad\;PSR}^\mathrm{s}$), single recycled pulsars ($f_\mathrm{rec\;PSR}^\mathrm{s}$), double-neutron-star (DNS) systems with at least one NS formed in an ECSN ($f_\mathrm{ECSN}^\mathrm{DNS}$) and NSs in binaries after the first SN ($f_\mathrm{BNS}^\mathrm{1st\;SN}$) in the considered toy population models.}
    \centering
    \begin{tabular}{lccccccc}
    \toprule 
    Model & $f_\mathrm{NS/BH}$ & $f_\mathrm{NS}^\mathrm{s}$ & $f_\mathrm{BH}^\mathrm{s}$ & $f_\mathrm{rad\;PSR}^\mathrm{s}$ & $f_\mathrm{rec\;PSR}^\mathrm{s}$ & $f_\mathrm{ECSN}^\mathrm{DNS}$ & $f_\mathrm{BNS}^\mathrm{1st\;SN}$ \\
    \midrule
    \midrule 
    Single & 4.3 & 86\% & 47\% & 91\% & 18\% & 32\% & 44\% \\
    Default & 14.7 & 80\% & 56\% & 88\% & 17\% & 65\% & 84\% \\
    CPS & 9.6 & 80\% & 58\% & 88\% & 17\% & 67\% & 80\% \\
    F12 rap.\ & 3.2 & 79\% & 36\% & 87\% & 17\% & 69\% & 62\% \\
    F12 del.\ & 1.9 & 75\% & 31\% & 84\% & 17\% & 75\% & 44\% \\    
    \bottomrule
    \end{tabular}
\end{table}

\begin{figure*}
    \begin{centering}
    \includegraphics{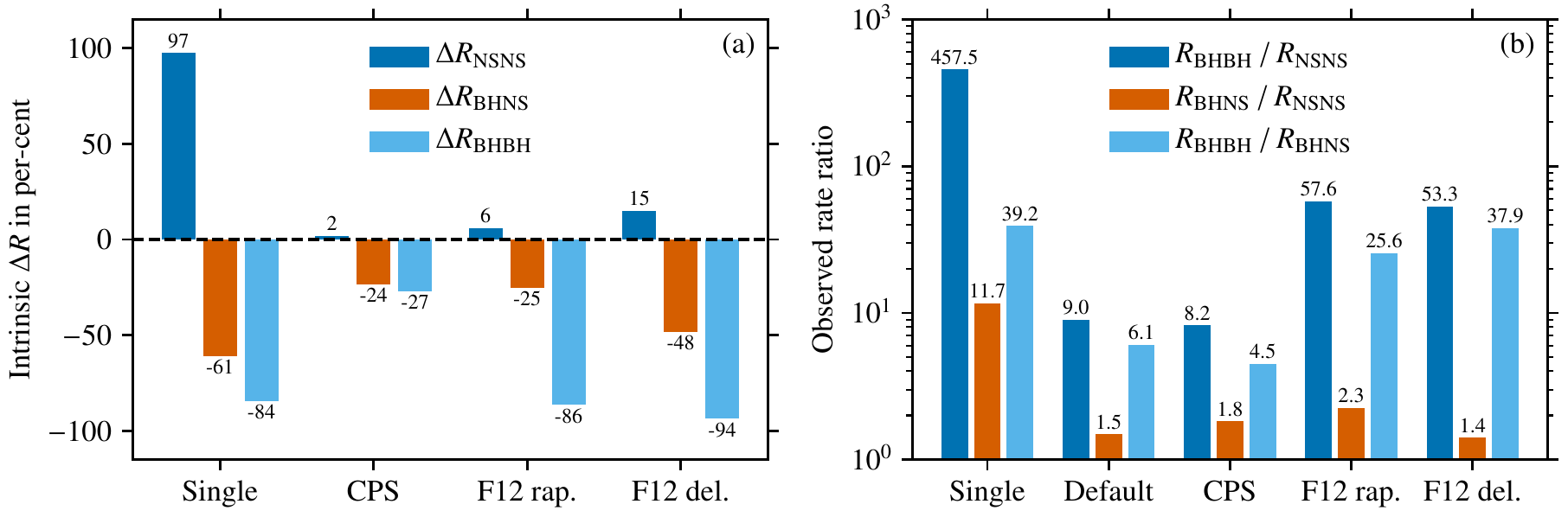} 
    \par\end{centering} \caption{Expected changes of intrinsic
    gravitational-wave merger rates $\Delta R$ of our default model with respect
    to the single, CPS, F12 rapid and F12 delayed models (panel a), and ratios
    of detection rates (panel b). The rate differences $\Delta R$ are defined
    such that positive (negative) values imply a higher (lower) rate in our
    default model compared to the other models.}
    \label{fig:merger-rates}
\end{figure*}

In all models, NSs are mostly single (${\gtrsim}80\%$ of all NSs) because of the
SN kicks, while about half of all BHs are single and the other half is in binary
systems with another compact object (Table~\ref{tab:pop-syn}). 

In particular, we consider the fraction of single radio pulsars to be a key
benchmark for SN kick prescriptions in population synthesis models. We define
NSs as radio pulsars if they are not expected to have accreted mass (\eg all
second-born NSs). Analogously, we denote NSs as recycled if they likely accreted
mass during their life (\eg the first-born NSs in binaries that are not broken
up by kicks). In all models, about 90\% of radio pulsars are single, which
appears to be in broad agreement with observations (J.\ Antoniadis and M.\
Kramer, private communication). About 15\%--20\% of all recycled pulsars are
single given our SN kick assumptions, which is an interesting prediction that
could be tested against future observations.

The NS-NS merger rate relates directly to double NS systems. In about 70\% of
these, at least one NS formed in an ECSN in our toy population (or more
generally in a low-kick SN; Table~\ref{tab:pop-syn}). This is because we assume
that binaries are not broken up by ECSNe. \citet{2017ApJ...846..170T} compile a
list of 13 double NS systems of which six systems have known component masses.
Four of these systems have one NS with a mass of $<1.3\,\msun$ and a rather
moderate eccentricity ($e<0.1\text{--}0.2$). From that we infer that roughly two
thirds (67\%) of known double NS systems have at least one NS that formed in a
low-kick SN (\eg an ECSN or from a low-mass iron core progenitor, hence the low
NS mass and moderate eccentricity). These statements are based on low-number
statistics and should not be over-interpreted, but provide credibility to our
assumed kick prescriptions and mass range for ECSNe from stripped binary stars. 

A direct consequence of these assumptions is that many if not most NS-NS mergers
in stripped binary populations are related to low-kick SNe and the mass range
over which they form. Binary break-up fractions and the size of the parameter
space from which low-kick SNe may be expected \citep[\eg
ECSNe;][]{2004ApJ...612.1044P, 2017ApJ...850..197P} are therefore probably a
significant source of uncertainty in the NS-NS merger rates of more elaborate
population synthesis models and warrants further investigation.

\subsection{Compact-object merger rates}\label{sec:merger-rates}

Here, we consider the intrinsic merger rates of compact objects in our toy
population model and also estimate detection rates. For the latter, we consider
the signal-to-noise $S/N$ of a merger event in a single gravitational-wave
detector. It is higher for a larger redshifted chirp mass, $M_\mathrm{chirp}
(1+z)$ (where $z$ is the redshift), and a closer (luminosity) distance to the
source, $d_\mathrm{L}$, \citep[\eg][]{1993PhRvD..47.2198F, 1996PhRvD..53.2878F},
\begin{equation}
    S/N \propto \frac{1}{d_\mathrm{L}}\left[ M_\mathrm{chirp} (1+z) \right]^{5/6}.
    \label{eq:signal-to-noise-gw-detector}
\end{equation}
The chirp mass $M_\mathrm{chirp}=\left(m_1 m_2\right)^{3/5}/\left(m_1 +
m_2\right)^{1/5}$ is directly measured from the frequency evolution of the
gravitational-wave signal and is larger for more massive component masses $m_1$
and $m_2$. Merger events with a larger chirp mass (\eg BH-BH mergers) therefore
result in a higher signal-to-noise and are observable over a larger
volume/fraction of the Universe. For a fixed signal-to-noise,
$d_\mathrm{L}\propto(M_\mathrm{chirp})^{5/6}$ such that the observable volume
$V_\mathrm{obs} \propto (d_\mathrm{L})^3 \propto (M_\mathrm{chirp})^{5/2}$. To
first order, we therefore expect that the rates of compact-object mergers
$R\propto(M_\mathrm{chirp})^{5/2}$ \citep[see also][]{1993PhRvD..47.2198F,
1996PhRvD..53.2878F}. Below, we use this scaling of the merger rates with chirp
mass to estimate the detection rates of gravitational-wave merger events (this
does not take into account the frequency dependence of the detector's
sensitivity curve).

Changes in the intrinsic merger rates of NS-NS, BH-NS and BH-BH with respect to
our default population model are shown in Fig.~\ref{fig:merger-rates}a ($\Delta
R_x = [R_\mathrm{default} - R_x]/R_x$, \ie positive [negative] differences are
for higher [lower] merger rates in the default model with respect to a
comparison model $x$). As expected from the NS to BH ratio
(Table~\ref{tab:pop-syn}), the NS-NS merger rate is intrinsically the highest in
our default model. When considering the detected NS-NS merger rate, the CPS
Model has a higher NS-NS merger rate than the default model, because the NS
masses and hence the chirp masses are on average smaller in the stripped binary
models than in the single star models employed in the CPS model.

It is not only the bare NS to BH ratio that sets the formation rate of NS-NS
mergers, but also the number of NSs that receive such kicks that binaries are
not broken up. Hence, low-kick SNe such as ECSNe can make a big difference: for
example, restricting the initial mass range for the occurrence of ECSNe in our
default model ($10.5\text{--}12.0\,\msun$) to the same as used in the
single-star population model ($9.5\text{--}10.0\,\msun$), would reduce the NS-NS
merger rate by about 60\% despite the now lower initial mass threshold for the
formation of NSs. This also explains a large part of the difference in the NS-NS
merger rate between our single and default population model. Also, the larger
intrinsic NS-NS merger rate of our default model in comparison to the CPS (2\%),
the F12 rapid (6\%) and the F12 delayed models (15\%) is due to the larger NS to
BH ratio (Table~\ref{tab:pop-syn}), as we apply the same assumptions on ECSNe in
these models.

\begin{figure*}
    \begin{centering}
    \includegraphics{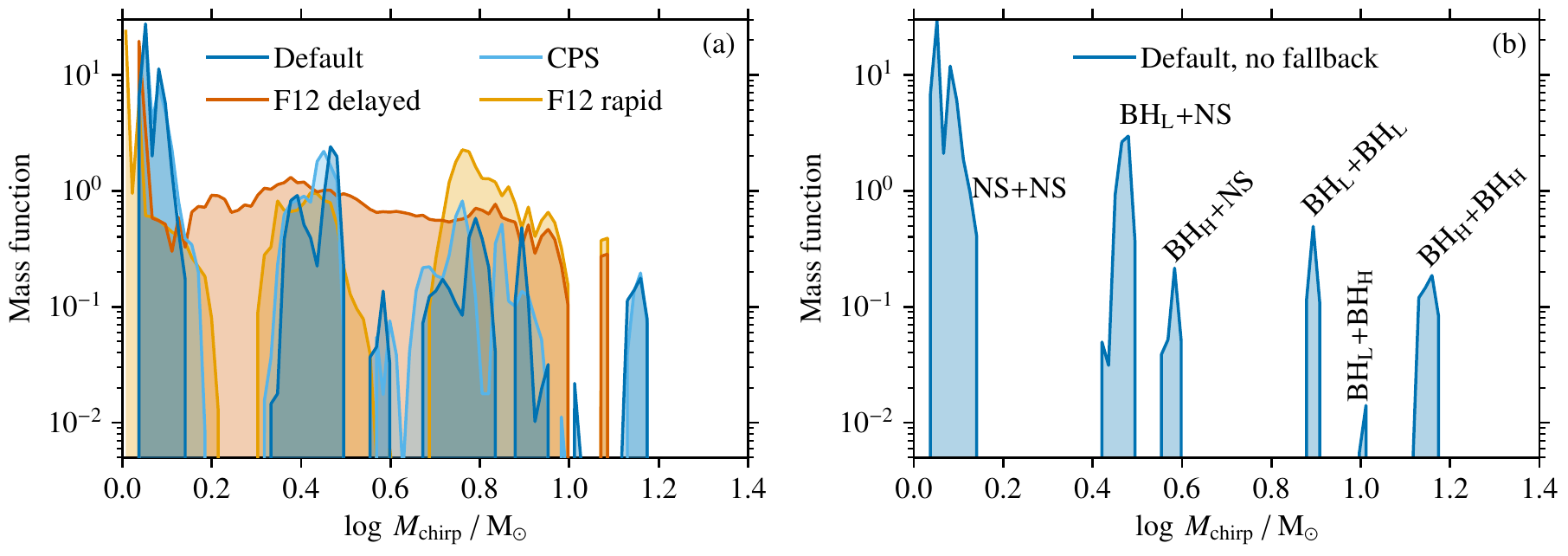} 
    \par\end{centering} \caption{Chirp-mass distributions of NS-NS, BH-NS and
    BH-BH mergers for our default, the CPS and the two F12 models (left panel
    a), and for the default model without fallback (right panel b).}
    \label{fig:chirp-mass-distribution}
\end{figure*}

The intrinsic and detected\footnote{Not shown in Fig.~\ref{fig:merger-rates}a.}
BH-NS and BH-BH merger rate is the lowest in our default model
(Fig.~\ref{fig:merger-rates}a). In particular the BH-BH merger rate can be lower
by even an order of magnitude ($\Delta R\approx -90\%$ with respect to the F12
rapid and delayed models) when considering that our stripped binary stars
produce NSs rather than BHs over a large initial-mass range
(Sect.~\ref{sec:compact-remnant-masses}). Even the seemingly small difference in
the mapping from CO core masses to NS or BH formation between our default and
the CPS model (\ie shifted compactness peak and BH formation at smaller
compactness in the single star models with respect to the stripped envelope
models) leads to a decrease in the intrinsic BH-NS and BH-BH merger rates by
about a quarter. We note again that our toy population lacks a potential
contribution from BH-BH mergers from stars evolving chemically homogenously in
close binaries and in dense stellar systems such as star clusters.

Within a differential analysis, we next consider the merger-rate ratios
$R_\mathrm{BHBH}/R_\mathrm{NSNS}$, $R_\mathrm{BHNS}/R_\mathrm{NSNS}$ and
$R_\mathrm{BHBH}/R_\mathrm{BHNS}$. These ratios are a promising way to compare
models to gravitational-wave observations, because the individual ratios are
influenced differently by certain physical mechanisms. Here, we focus on the
question of how the pre-SN structures relate to NS and BH formation. The
single star model clearly stands out with the highest rate ratios. The merger
rate ratios are so high because of the relatively low NS to BH ratio
(Table~\ref{tab:pop-syn}), the limited initial-mass range of NS formation in
low-kick SNe (here ECSNe) and the on average larger BH masses of the single-star
models compared to the stripped-binary models. The latter leads to larger chirp
masses and thus higher detection rates. 

In the F12 rapid and delayed models, the individual BH masses are smaller than
those in the single star models because of envelope stripping. Hence, also the
chirp masses of mergers involving BHs are smaller and thereby the detected BH-BH
and BH-NS merger rate. At the same time, the NS-NS rate is increased due to the
larger contribution of NSs from low-kick SNe (here ECSNe). Together this
drastically reduces the $R_\mathrm{BHBH}/R_\mathrm{NSNS}$ and
$R_\mathrm{BHNS}/R_\mathrm{NSNS}$ merger-rate ratios
(Fig.~\ref{fig:merger-rates}b). The ratio $R_\mathrm{BHBH}/R_\mathrm{BHNS}$
stays about the same, because the NS to BH ratios are quite similar in the three
models (Table~\ref{tab:pop-syn}).

Comparing the F12 rapid and delayed models to our default and CPS models, the
initial-mass range for NS formation is larger in the latter models such that
fewer BHs are formed. As shown in Fig.~\ref{fig:merger-rates}a, this decreases
the intrinsic BH-BH and BH-NS merger rates and increases the NS-NS merger rate.
Consequently, $R_\mathrm{BHBH}/R_\mathrm{NSNS}$ and
$R_\mathrm{BHBH}/R_\mathrm{BHNS}$ decrease and $R_\mathrm{BHNS}/R_\mathrm{NSNS}$
remains about the same in our toy population model
(Fig.~\ref{fig:merger-rates}b). To fully understand this picture, we note that
the F12 rapid and delayed models predict more lower-mass BHs than our default
and CPS models such that they contribute relatively less in the detected merger
rates involving BHs. For example, this explains why the intrinsic BH-BH merger
rate is almost a factor of 10 higher in the F12 models than in our default model
but the detected $R_\mathrm{BHBH}/R_\mathrm{NSNS}$ ratio is only a factor of
${\approx}\,6$ larger.

\subsection{Chirp-mass distribution}\label{sec:chirp-mass-distribution}

In the following, we compute chirp-mass distributions of NS-NS, BH-NS and BH-BH
mergers from their detection rates\footnote{This means that the intrinsic
chirp-mass distribution can be obtained by dividing the shown distribution by
$M_\mathrm{chirp}^{5/2}$ (see Sect.~\ref{sec:merger-rates}).}. In particular, we
will highlight characteristic features that can be directly linked to our models
of stripped binary stars. 

The chirp-mass distribution of the default, CPS and the F12 models are shown in
Fig.~\ref{fig:chirp-mass-distribution}a. In
Fig.~\ref{fig:chirp-mass-distribution}b, we only show results for the default
model with switched-off fallback to highlight the most important features. The
chirp-mass distribution is multi-modal and there are three main components from
NS-NS, BH-NS and BH-BH mergers at $\log M_\mathrm{chirp}\, {\approx}\, 0.1$,
${\approx}\,0.5$ and ${\approx}\,1.0$ that are further split into sub-components
(Fig.~\ref{fig:chirp-mass-distribution}b). To help understand these features, we
recall that the compact object masses from our stripped binary models without
fallback (stripped stars contribution in top panel of
Fig.~\ref{fig:comf-default-and-fallback-model}) are essentially split into three
main groups: NSs (${\approx}\,1.4\,\msun$), lower-mass BHs (BH\textsubscript{L};
${\approx}\,9\,\msun$) from stars in the compactness peak and higher-mass BHs
(BH\textsubscript{H}; ${\approx}\,17\,\msun$). Mergers of compact objects from
these three groups then explain the six components observed in
Fig.~\ref{fig:chirp-mass-distribution}b: 
\begin{itemize}
    \item the NS-NS contribution at $\log M_\mathrm{chirp}\, {\approx}\, 0.09$,
    \item the two BH-NS contributions at $\log M_\mathrm{chirp}\, {\approx}\,
    0.46$ (BH\textsubscript{L}+NS) and $\log M_\mathrm{chirp}\, {\approx}\,
    0.57$ (BH\textsubscript{H}+NS) and 
    \item the three BH-BH contributions at $\log M_\mathrm{chirp}\, {\approx}\,
    0.89$ (BH\textsubscript{L}+BH\textsubscript{L}), $\log M_\mathrm{chirp}\,
    {\approx}\, 1.03$ (BH\textsubscript{L}+BH\textsubscript{H}) and $\log
    M_\mathrm{chirp}\, {\approx}\, 1.17$
    (BH\textsubscript{H}+BH\textsubscript{H}). 
\end{itemize}
Also the contribution from low-mass NSs from ECSNe of $1.25\,\msun$ are visible
by the double peak in the chirp-mass distribution of the NS-NS mergers and the
shoulders towards lower chirp masses of the two BH+NS contributions. 

Accounting for fallback in the default model
(Fig.~\ref{fig:chirp-mass-distribution}a), we see that further components are
added to the chirp-mass distribution. Assuming an average fallback BH mass of
about $6\,\msun$ (BH\textsubscript{F}), adds four components at 
\begin{itemize}
    \item $\log M_\mathrm{chirp}\, {\approx}\,0.38$ (BH\textsubscript{F}+NS), 
    \item $\log M_\mathrm{chirp}\, {\approx}\, 0.70$
    (BH\textsubscript{F}+BH\textsubscript{F}), 
    \item $\log M_\mathrm{chirp}\, {\approx}\, 0.80$
    (BH\textsubscript{F}+BH\textsubscript{L}) and 
    \item $\log M_\mathrm{chirp}\, {\approx}\, 0.93$
    (BH\textsubscript{F}+BH\textsubscript{H}).
\end{itemize}
Overall this leads to a smearing out of the main features.

Comparing the CPS model to the default model, we see that some features are
shifted to lower chirp masses. For example, this is because the compactness peak
is at lower CO core masses in single stars, so the BH masses from stars in the
compactness peak of the CPS model is at lower masses than that of the default
model (\cf Fig.~\ref{fig:com-toy-models}). All components in the chirp-mass
distribution that involve BH\textsubscript{L} are thus slightly shifted to lower
masses. The F12 rapid model also shows clearly separated contributions, but with
more weight on BH-BH mergers compared to the other models, in particular the
default and CPS models. The BH masses span a larger mass range in the F12 rapid
model than in the default and CPS models, which leads to less substructured and
broader BH-NS and BH-BH components in the chirp mass distribution. The most
dramatic difference in the chirp-mass distribution is found in the F12 delayed
model. In this model, there is no mass gap between NSs and BHs. So there are no
longer individual components in the NS--BH mass distribution and thus also not
in the chirp-mass distribution.

Our toy populations are only for stars of one metallicity and we have not
incorporated the cosmic star-formation history. The chirp-mass distribution that
will be available from gravitational-wave observations is going to be a
convolution of various populations of different metallicities. As discussed in
Sect.~\ref{sec:compact-remnant-masses}, in particular the BH masses, \eg, depend
on the total (wind) mass loss of stars. Stronger (weaker) winds at higher
(lower) metallicities tend to shift BH masses to lower (higher) values. This
way, we expect the components in the chirp-mass distribution to be effectively
further smeared out in real observations.

%
%
\section{Discussion}\label{sec:discussion}

\subsection{General remarks and uncertainties}\label{sec:uncertainties}

Wind mass-loss rates of massive stars are uncertain
\citep[\eg][]{2014ARA&A..52..487S}. This is true for winds of main-sequence,
(red) supergiant and WR stars, and such uncertainties directly translate to
various stellar properties up to core collapse
\citep[\eg][]{2017A&A...603A.118R}. The latter mass-loss rates are of particular
relevance for stripped stars that become WR stars
\citep[\eg][]{2002ApJ...578..335F}, and extrapolating wind mass-loss rates from
classical WR stars to lower mass helium stars after binary envelope stripping is
problematic \citep[][]{2017A&A...607L...8V, 2019MNRAS.486.4451G}. Stronger winds
have the tendency to not only reduce the overall stellar masses but also core
masses. Similarly to the most extreme form of mass loss, namely binary envelope
stripping, the more mass is lost by winds, the larger is the core carbon mass
fraction at core helium depletion with similar consequences as discussed in this
work for \casea and~B stripped stars. Enhanced wind mass loss after core helium
burning will most likely not drastically affect the pre-SN structure,
analogously to what is found in \casec envelope stripping. While the final fate
(\eg explodability, compact remnant mass, explosion energy etc.) is a
non-monotonic function of the CO core mass, the general expectation is that
stronger winds make it easier for stars to explode, enhance the NS formation
rate, lead to more energetic explosions, faster kicks and higher nickel yields,
and vice versa for weaker winds. Stellar winds scale with metallicity (\cf
Sect.~\ref{sec:single-star-models}), such that stars at higher metallicity than
studied here experience more mass loss with the above described consequences
(and vice versa for stars at lower metallicity, hence weaker winds). 

LBV-like mass loss via steady winds and eruptive episodes is not considered
here, but will contribute to the total mass loss of stars. As discussed in
Sect.~\ref{sec:ns-bh-mass-distr-default} and shown in
Fig.~\ref{fig:compact-remnant-masses}, this is most relevant for single stars
that may produce BHs. Stripped binary stars only cross the LBV instability strip
in the HR diagram before envelope removal and usually do not reach this regime
thereafter. The enhanced mass loss through LBV-like winds will reduce the final
total mass of stars and hence the BH mass. The BH masses from single stars
reported in this work are therefore rather upper limits (\cf
Fig.~\ref{fig:compact-remnant-masses}).

In this study, we have assumed that BHs form by direct collapse, \ie without
neutrino emission. However, if BH-forming progenitors first form a proto-NS that
then subsequently collapses to a BH, neutrinos can radiate away a non-negligible
amount of energy and hence mass (${\approx}0.3\,\msun$) such that the outer
layers of a star might get unbound and power a weak transient
\citep{1980Ap&SS..69..115N, 2013ApJ...769..109L, 2018MNRAS.477.1225C}. This in
turn effectively reduces the BH mass. Because of this, it is often assumed that
the resulting BH mass is 90\% of the final baryonic mass of a star
\citep[\eg][]{2012ApJ...749...91F}. If most BHs do not form by direct
collapse, our BH masses may be overestimated by about 10\%. This most likely
applies to the BHs formed by fallback in our models.

Convective (core) overshooting or more generally speaking convective boundary
mixing greatly affects stars and remains an active field of research
\citep[\eg][]{2012sse..book.....K, 2015A&A...575A.117S}. Convective core
overshooting driven by the major burning cycles enlarges core masses and
prolongs the respective burning cycle. For example, on the main sequence and
during core helium burning, this implies that overshooting sets the relation
between initial and CO/iron core mass, and hence, \eg, the relative numbers of
NSs and BHs formed by a stellar population. More directly, a larger core
overshooting on the main sequence allows stars to evolve to cooler temperatures,
which has further implicit consequences: cooler stars lose more mass via winds
and the total mass loss then affects the core evolution and hence final fate. We
include moderate ``step overshooting'' of 0.2 pressure scale heights for core
hydrogen and helium burning. A similar amount of overshooting is applied in some
state-of-the-art stellar grids \citep[\eg][]{2012A&A...537A.146E,
2015MNRAS.452.1068C}, whereas in others an even larger overshooting is applied
to match the wide main sequence at masses ${\gtrsim}\,10\text{--}15\,\msun$
\citep[\eg][]{2011A&A...530A.115B, 2014A&A...570L..13C, 2016ApJ...823..102C}.
Also, during the late burning stages, convective boundary mixing in shell burning
regions may trigger shell mergers, which could result in enhanced nuclear
burning with consequences for the stellar structure. We do not consider
convective boundary mixing beyond core helium burning.

The ${}^{12}\mathrm{C}(\alpha,\gamma){}^{16}\mathrm{O}$ reaction is key to
several stellar astrophysical applications, but remains uncertain
\citep[\eg][]{2017RvMP...89c5007D}. In Sect.~\ref{sec:pre-sn-structure}, we
stress the importance of the central carbon abundance after core helium burning
for the evolution of stars towards core collapse. The
${}^{12}\mathrm{C}(\alpha,\gamma){}^{16}\mathrm{O}$ is key as it sets the rate
at which carbon is converted into oxygen during the later phases of helium
burning and thus sets the carbon abundance after this burning episode. Moreover,
depending on the strength of this rate, the energy release and hence size of the
convective core during helium burning could even be larger or smaller. As shown
by, \eg, \citet{2001ApJ...558..903I}, the exact
${}^{12}\mathrm{C}(\alpha,\gamma){}^{16}\mathrm{O}$ rate may influence whether
core carbon burning proceeds convectively or radiatively with all the
consequences for the pre-SN structure and SN as also discussed here. For
example, the CO core mass range corresponding to the compactness peak and hence
a possible island of BH formation is intimately connected to this reaction rate
\citep[see also][]{2020MNRAS.492.2578S}. However, regardless of the value of
this rate, \casea and~B stripped binary stars will always give rise to stars
with a systematically larger carbon abundance after core helium burning than
single and \casec stripped stars (Fig.~\ref{fig:core-c-mass-fraction}).

Our models employ \mesa's approximate \texttt{approx21\_cr60\_plus\_co56.net}
nuclear reaction network. Such approximate networks are limited in their ability
to accurately obtain the electron fraction $Y_\mathrm{e}$ after oxygen burning.
Larger networks are thus required to follow more accurately the neutronization
and subsequent core collapse upon approaching the finite-temperature, effective
Chandrasekhar mass \citep[\eg][]{1978ApJ...225.1021W, 2016ApJ...821...38S,
2016ApJS..227...22F, 2018ApJ...860...93S}. For example,
\citet{2018ApJ...860...93S} find variations in the central electron fraction at
core collapse of $Y_\mathrm{e}=0.433\text{--}0.438$ and
$Y_\mathrm{e}=0.438\text{--}0.444$ in an intially $15\,\msun$ and $25\,\msun$
star when switching from the ``workhorse'' 19 isotope network to a large network
with about 300 isotopes that is fully coupled to the solver of the stellar
structure equations. For the same initial masses, \citet{2016ApJS..227...22F}
report variations of $Y_\mathrm{e}\approx0.415\text{--}0.425$ and
$Y_\mathrm{e}\approx0.438\text{--}0.440$, respectively, in their figure 21. In
the \citeauthor{2018ApJ...860...93S} models, these variations correspond to
final compactness values of $\xi_{2.5}=0.139\text{--}0.187$ and
$\xi_{2.5}=0.240\text{--}0.311$ for the $15\,\msun$ and $25\,\msun$ stars,
respectively. While such variations are non-negligible and add to the overall
uncertainty on the final pre-SN structures of massive stars, we are here more
interested in the systematic differences between genuine single and stripped
binary stars.

As studied extensively by \citet{2016ApJS..227...22F}, also the numerical
meshing affects pre-SN structures, and time-stepping and other numerical solver
choices will do so, too (\eg, we use \texttt{varcontrol\_target = 1d-4}). For
physical convergence, \citet{2016ApJS..227...22F} recommend the use of zones no
bigger than $0.01\,\msun$ in \mesa. While the absolute zoning is clearly
important, adaptive mesh refinement and relative zoning are also crucial. To
this end, the spatial zoning in our models is limited to relative cell masses of
at most $10^{-3}$ (\texttt{max\_dq = 1d-3}) and cells can be smaller (\eg,
\texttt{mesh\_delta\_coeff = 0.6} in our models).

\subsection{Comparison to other stellar evolution models}\label{sec:other-stellar-models}

Several groups have computed pre-SN stellar models \citep[\eg][to mention a few
more recent publications]{2002RvMP...74.1015W, 2004A&A...425..649H,
2007PhR...442..269W, 2013ApJ...764...21C, 2014ApJ...783...10S,
2016ApJ...821...38S, 2016ApJS..227...22F, 2018ApJ...860...93S,
2020ApJ...890...43C}. All of these models have in common that they predict a
steep increase in the core compactness at the transition from convective to
radiative core carbon burning. This transition and compactness peak is at
different masses in the models, because of different physics assumptions. For
example, convective core overshooting, the still uncertain
${}^{12}\mathrm{C}(\alpha,\gamma){}^{16}\mathrm{O}$ nuclear reaction rate, and
wind mass loss all affect the location of the compactness peak \citep[see
also][]{2001NewA....6..457B, 2020MNRAS.492.2578S}.

Furthermore, the compactness is observed to increase again beyond the
compactness peak in many models, but this increase is closely connected to the
adopted wind mass-loss routines, in particular the WR mass-loss rates
\citep[\eg][]{1993ApJ...411..823W, 2002ApJ...578..335F}. Larger wind mass loss
(and in the most extreme case envelope stripping in binaries) are seen to reduce
the compactness \citep[see also][]{2018ApJ...860...93S}, just as is found in
this work. A systematic study looking into the evolution of pure CO cores with
different initial carbon abundances towards core collapse
\citep{2020arXiv200503055P} recovers this general landscape of compactness (or
alternatively central entropy or iron core mass, \cf
Sect.~\ref{sec:pre-sn-structure}). Hence, this appears to be robust across
different codes and assumptions on key physical processes.

Binary envelope stripping is often mimicked by considering helium star evolution
\citep[\eg][]{1995ApJ...448..315W, 2016MNRAS.459.1505M, 2019ApJ...878...49W},
but full binary evolution models up to core collapse are also available
\citep[\eg][]{1999A&A...350..148W, 2002ApJ...578..335F}. From the evolution of
stars without a hydrogen-rich envelope, \citet{1996ApJ...457..834T} and
\citet{2001NewA....6..457B} realized that stripped stars have a different pre-SN
core structure and that this is related to the higher carbon mass fraction in
stripped stars compared to single stars \citep[see also][and
Sect.~\ref{sec:pre-sn-structure}]{2004ApJ...612.1044P}. As we have shown in this
work, these genuinely different core structures of stripped stars lead to a
compactness peak shifted to higher CO core masses and a compactness that
increases only at higher CO core masses in stripped stars compared to single
stars (Fig.~\ref{fig:compactness-and-central-entropy}). This is in line with,
\eg, the findings of \citet{2019ApJ...878...49W} from helium star models
\citep[see also][]{2020ApJ...890...51E}.

Our results further suggest that \casec envelope removal, \ie envelope stripping
after core helium burning (Fig.~\ref{fig:definition-mt-cases}), leads to similar
compactness values as in single stars (\eg the compactness peak is at the same
CO core masses, Fig.~\ref{fig:compactness-and-central-entropy}). This is because
the core structure in terms of CO core mass and core carbon mass fraction is
already set at the end of core helium burning. However, we find that the
explosion energies, kick velocities and also compact remnants masses of \casec
binary models are not necessarily the same as in single stars. For example, the
binding energy of the envelope will be different as is the helium core mass.
Helium star models are thus less suitable to describe \casec envelope removal.

\subsection{NS and BH formation}\label{sec:ns-bh-formation}

The transition from NS to BH formation is sometimes thought to occur at a
critical initial mass \citep[see \eg the well known schematic picture
in][]{2003ApJ...591..288H}. Such a simple picture is not supported by more
recent studies and a more complex, non-monotonic pattern of NS and BH formation
has emerged \citep[\eg][]{2016ApJ...821...38S, 2016ApJ...818..124E}. It has
even been discussed whether these patterns are stochastic in nature
\citep{2015ApJ...799..190C}.

Here, we find a somewhat less random landscape of NS and BH formation that
closely follows the final compactness (or alternatively iron core mass or
central entropy, see Sect.~\ref{sec:pre-sn-structure}). Regardless of whether
the pre-SN star is a genuine single star or has been stripped off its envelope
by binary mass exchange, all models give rise to a compactness peak related to
the transition from convective to radiative core carbon burning for which BH
formation is found in our models. Beyond the compactness peak, NS formation is
possible again, but we also find failed SNe and BH formation by fallback in
cases where an explosion is triggered initially, but stalls later on. In even
more massive stars, a SN explosion is not found and BHs are formed
(Fig.~\ref{fig:compact-remnant-masses}).

Observationally, BH rather than NS formation may manifest itself in the
highly-debated missing red-supergiant (RSG) problem, \ie in possibly missing
high-luminosity stars ($\log L/\lsun \gtrsim 5.1$) among observed progenitors
of Type IIP SNe \citep[\eg][]{2009ARA&A..47...63S, 2015PASA...32...16S}. This
luminosity threshold contrasts with the inferred maximum luminosity of RSGs of
$\log L/\lsun \approx 5.5$ in the Milky Way, and Small and Large Magellanic
Clouds \citep{2018MNRAS.478.3138D}, Hence, RSGs with $\log L/\lsun \approx
5.1\text{--}5.5$ may not explode, but rather collapse to a BH. It should be
noted that obtaining robust bolometric luminosities from pre-explosion
photometry is challenging \citep[\eg][]{2018MNRAS.474.2116D} and the maximum
luminosity of SN~IIP progenitors may be higher by
$0.1\text{--}0.2\,\mathrm{dex}$ \citep[\eg][]{2020MNRAS.493..468D}. Still, the
missing RSG problem appears to persist at the $1\text{--}2\sigma$ significance
level \citep[][]{2020MNRAS.493..468D}.

By monitoring a million supergiants for failed SNe \citep{2008ApJ...684.1336K},
a first RSG of $\log L/\lsun\approx 5.3\text{--}5.5$ has been found to disappear
without a SN~IIP explosion \citep[N6946-BH1;][]{2015MNRAS.450.3289G,
2017MNRAS.468.4968A, 2020arXiv200715658B}. Instead, the star gave rise to a weak
transient similar to the failed SN models of a star collapsing to a BH by
\citet{2013ApJ...769..109L}.

Taken together, it seems that some RSGs collapse to a BH. Our single star models
in the compactness peak ($M_\mathrm{ini}\approx21.5\text{--}23.5\,\msun$) reach
luminosities of $\log L/\lsun\approx 5.3\text{--}5.5$ at core collapse and are
predicted to form BHs (our stripped envelope models will give rise to
SN~Ib/c). This association of BH formation and the high compactness in RSGs
that burn carbon radiatively in their core has also been made by
\citet{2020MNRAS.492.2578S}. They also find that the
${}^{12}\mathrm{C}(\alpha,\gamma){}^{16}\mathrm{O}$ reaction rate leads to
significant uncertainty regarding the question of which RSG collapse to a BH
(see discussion above). 

Our single star models predict NS formation for stars beyond the compactness
peak such that it is conceivable that somewhat more luminous RSGs ($\log
L/\lsun\gtrsim 5.4\text{--}5.5$ in our single star models) explode again in Type
IIP SNe and produce NSs. However, stellar winds in these stars are strong (see
also Sect.~\ref{sec:uncertainties}) and the associated SN explosion might
transition to Type II-L, II-b or even I-b, depending on the remaining
hydrogen-rich envelope. 

The initial mass range for NS formation is considerably larger in our stripped
binary models than in the single stars (Table~\ref{tab:mini-nss}). This has
important ramifications for, \eg, the occurrence of NSs and BHs in young star
clusters and in massive binaries. Understanding the existence of the X-ray
pulsar CXO\,J164710.2-455216 in the young star cluster Westerlund~1 with a
turnoff mass of about $40\,\msun$ is difficult if not impossible from a
single-star point of view \citep[][]{2006ApJ...636L..41M}, because single stars
of initially $>40\,\msun$ are expected to form BHs. This implies that the
progenitor star of CXO\,J164710.2-455216 experienced strong mass loss, \eg, via
binary envelope stripping or an LBV-like (giant) eruption \citep[see also][for a
binary-stripping interpretation]{2008ApJ...685..400B}. 

Similarly, the high present-day masses of companions to NSs in some X-ray
binaries may only be understood when accounting for the binary-mass stripping
history of the NS progenitors. However, this depends on how conservative mass
transfer was in the evolution leading to the X-ray binary. For example, in the
X-ray binary Wray 977/GX 301-2 (4U 1223-62), the B hypergiant companion Wray 977
has a current day inferred mass of $39\text{--}53\,\msun$
\citep{2006A&A...457..595K}, \ie the total initial mass of the binary must have
been larger than that. Within a non-conservative mass-transfer scenario,
\citet{1998A&A...331L..29E} suggest an initial progenitor mass of $>50\,\msun$
for the X-ray pulsar GX 301-2, while \citet{1999A&A...350..148W} show that,
within a more conservative mass-transfer scenario, an initial progenitor mass of
$26\,\msun$ suffices to explain the present-day configuration of Wray 977. In
both scenarios, our stripped binary models indeed predict NS formation.

\subsection{NS--BH mass gap and X-ray binaries}\label{sec:mass-gap}

Depending on the amount of fallback, our models do or do not show a mass gap
between NSs and BHs (Sect.~\ref{sec:ns-bh-mass-distr} and
Fig.~\ref{fig:comf-default-and-fallback-model}). Without fallback, the gap is at
${\approx}\,2\text{--}9\,\msun$ in the stripped binary stars and it narrows down
to ${\approx}2\text{--}5\,\msun$ in the 50\% fallback models. For smaller
fallback fractions, the gap narrows down further. Hence, the extent and
existence of the gap is set by the SN physics. Only for sufficiently higher
total mass loss than assumed here, models would predict pre-SN stellar masses
that could lead to a rather smooth transition from NSs to BHs without a gap
\citep[see also][]{2002ApJ...578..335F}. This could be realized in stars at a
metallicity significantly higher than solar. 

Models of neutrino-driven explosions coupled to stellar models are in agreement
with our conclusions on a possible NS--BH mass gap \citep{2012ApJ...757...69U,
2015ApJ...801...90P, 2016ApJ...821...38S, 2016MNRAS.460..742M,
2018ApJ...860...93S}. Other explosion models do not predict such a gap
\citep[\eg][]{2001ApJ...554..548F, 2008ApJ...679..639Z, 2012ApJ...749...91F},
but this depends on the assumption of whether an explosion is triggered rapidly
or delayed. In the latter case (\cf F12 delayed model in
Fig.~\ref{fig:chirp-mass-distribution}), accretion of fallback material on the
central compact remnant closes the NS--BH mass gap that is predicted in the
rapid explosion model. As before, we thus conclude that fallback or accretion
onto a central compact object in delayed SN explosions is the main mechanism
that governs the existence and width of a possible NS--BH mass gap \citep[see
also][]{2012ApJ...757...91B, 2014ApJ...785...28K, 2020MNRAS.495.3751C}.

From observations of Galactic X-ray binaries, \citet{2010ApJ...725.1918O} and
\citet{2011ApJ...741..103F} infer BH masses of ${\gtrsim}\,5\,\msun$ \citep[see
also][]{2017hsn..book.1499C}. Together with a maximum NS mass of about
$2\,\msun$ \citep[\eg][]{2010Natur.467.1081D, 2013Sci...340..448A}, this
suggested a NS--BH mass gap in the range $2\text{--}5\,\msun$.
\citet{2012ApJ...757...36K} caution that systematics could have lead to
overestimated BH masses from some X-ray binaries, which may reduce the
determined BH masses to a value inside the mass gap. 

The existence of this gap is challenged. First, \citet{2019Sci...366..637T}
report the discovery of a ${\approx}\,83\,\mathrm{d}$ binary containing a
rapidly rotating giant star and a $3.3^{+2.8}_{-0.7}\,\msun$ dark object, most
likely a BH. Secondly, there are claims with potentially large systematic
uncertainties of pulsars with masses above $2\,\msun$, \eg the
${\approx}\,2.7\,\msun$ millisecond pulsar in the globular cluster NGC\,6440
\citep{2008ApJ...675..670F}. Thirdly, in gravitational microlensing events,
compact objects in the NS--BH mass gap may have been found
\citep{2020A&A...636A..20W}. The individual mass measurements of these objects
are consistent with masses ${\lesssim}\,2\,\msun$ and ${\gtrsim}\,5\,\msun$
within their uncertainties. However, interpreting all individual observations
simultaneously within a certain model for the compact-object mass distribution,
\citet{2020A&A...636A..20W} strongly disfavour the existence of a NS--BH mass
gap at $2\text{--}5\,\msun$. Fourthly, from gravitational-wave emission of
merging compact objects, \citet{2020ApJ...896L..44A} report the merger of a
$22.2\text{--}24.3\,\msun$ BH with a $2.50\text{--}2.67\,\msun$ compact
companion (GW190814; 90\% credibility intervals). Previously, only BHs of masses
of ${\gtrsim}\,7\,\msun$ have been observed in gravitational-wave merger events
\citep[\eg][]{2019PhRvX...9c1040A}. The previous BH-BH mergers did not invoke
BHs of such unequal masses, which may point to a rarer formation channel. It
thus seems that the previous suggestion of the existence of a NS--BH mass gap
may no longer hold true.

In principle, compact objects in X-ray binaries can trace intrinsically
different compact-object populations, may have different evolutionary paths and
even different origins than compact objects in wider binaries without X-ray
emission, microlensing surveys and gravitational-wave merger events. For
example, microlensing can find the remnants of NS-NS mergers \citep[\eg the
$2.7\,\msun$ remnant of GW170817;][]{2017PhRvL.119p1101A}, primordial BHs
\citep[\eg][]{1967SvA....10..602Z, 1971MNRAS.152...75H, 2016PhRvD..94h3504C} and
also BHs ejected from binary systems by SN kicks
\citep[\eg][]{2005ApJ...625..324W, 2009ApJ...697.1057F}. Similarly, the
unusually large mass ratio of GW190814 may point to a rare formation channel
\citep{2020ApJ...896L..44A}.

If BHs form via direct collapse, they might not receive a kick. In our fallback
cases, as discussed in Sect.~\ref{sec:kicks}, BHs receive a kick and the kick
velocity is larger for fewer fallback material. This means that the least
massive BHs formed via fallback receive the largest kicks and the most massive
BHs receive the smallest kicks. In our models, BHs $\lesssim 9\,\msun$ in
stripped binary stars (and $\lesssim 12\,\msun$ in single stars) receive kicks
whereas more massive BHs might not. From our toy population model we see that
about 50--60\% of BHs are single and no longer in binaries. Within our models,
the formation of BHs with the least fallback is most likely to break up binaries
by SN kicks and this population of low-mass BHs may explain some of the NS--BH
mass gap objects found by \citet{2020A&A...636A..20W}. This idea does not
exclude the existence of BHs of $2\text{--}5\,\msun$ in X-ray binaries, but it
would make it less likely and could help to reduce some of the apparent tension.

From observations of X-ray binaries, it is clear that some BHs formed in a SN
and received a kick, because the companion stars show chemical signatures of a
SN event and the binary has a high space velocity \citep[see \eg GRO\,J1655-40
also known as Nova Sco 1994;][]{1995MNRAS.277L..35B, 1999Natur.401..142I,
2002ApJ...567..491P, 2005ApJ...625..324W}. But also the large distance to the
plane of the Milky Way of BH X-ray binaries suggests that they have received a
kick \citep[\eg][]{2004MNRAS.354..355J, 2015MNRAS.453.3341R}. As noted by
\citet{2017hsn..book.1499C}, it is interesting that Cygnus\,X-1 and
GRS\,1915+105, X-ray binaries with BHs of ${\approx}\,15\,\msun$ and
${\approx}\,12\,\msun$, respectively, likely received no or only a small kick
\citep[\eg][]{2003Sci...300.1119M, 2014ApJ...796....2R}, whereas XTE\,J1118+480
\citep[\eg][]{2001Natur.413..139M,2009ApJ...697.1057F}, the aforementioned
GRO\,J1655-40 and likely also V404 Cygni \citep[\eg][]{2009ApJ...706L.230M,
2009MNRAS.394.1440M} with BH masses of ${\approx}\,7\text{--}8\,\msun$,
${\approx}\,5.4\,\msun$ and ${\approx}\,9\,\msun$, respectively, probably formed
with a natal kick. This could indeed be a hint that less massive BHs are formed
with a higher kick than their more massive counterparts.

The most massive BH in our stripped binary models is ${\approx}\,20\,\msun$, \ie
somewhat larger than one of the most massive BHs observed in an X-ray binary in
the Milky Way \citep[\ie Cyg\,X-1 with
${\approx}15\,\msun$,][]{2011ApJ...742...84O}. As discussed in
Sect.~\ref{sec:uncertainties}, if BHs do not form by direct collapse, but if a
proto-NS is formed first that can emit neutrinos, our BH masses are somewhat
overestimated. In general, the most massive BH in stripped stars is closely
linked to the WR wind mass loss after the loss of the hydrogen-rich envelope
\citep[see also][]{2002ApJ...578..335F}. So while the difference between our
maximum BH mass and that of Cyg X-1 is not large, it may suggest that the WR
wind mass loss applied in our models is too low.

Starting from the observed WR+OB binary stars in the Galaxy,
\citet{2020A&A...636A..99V} predict the expected population of wind-fed BH
high-mass X-ray binaries. Under the assumption that the WR stars collapse to BHs
with kicks insufficient to break up the WR+OB binaries, they expect to find more
than 100 such BH X-ray binaries whereas only one (Cyg X-1) is observed. Given
this stark discrepancy, \citeauthor{2020A&A...636A..99V} conclude that the WR
stars either formed BHs that received a significant kick or that they collapsed
to NSs. Most of the Galactic WR+OB binaries considered in
\citet{2020A&A...636A..99V} have so short orbital periods that they must have
undergone a past mass transfer episode in which the current WR star lost its
envelope to its companion star. From our models, we thus expect that most of
these stripped WR stars will form a NS at the end of their life, thereby
offering a natural solution to this discrepancy.

\subsection{Explosion energies, nickel yields and kicks}\label{sec:discussion-explosion-energy-nickel-kicks}

Observations of SN light curves suggest that, on average, stripped-envelope SNe
produce more ${}^{56}\, \mathrm{Ni}$ and are more energetic than SN IIP
\citep[\eg][]{2019A&A...628A...7A, 2020A&A...641A.177M, 2020MNRAS.496.4517S,
2020arXiv200906683A} -- just as predicted by our models
(Sect.~\ref{sec:explosion-energy-and-nickel-mass}). At first glance, the
observed differences are quite dramatic, for example the recent meta-analysis of
\citet{2019A&A...628A...7A} gives median values of $0.032\,\msun$ for the nickel
mass $M_\mathrm{Ni}$ for Type~II SNe and $0.16\,\msun$ for Type~Ib/c SNe
(excluding broad-lined Type~Ic events). Closer inspection suggests less dramatic
differences because of selection effects and systematic uncertainties in
determining $M_\mathrm{Ni}$.

For Type~IIP SNe, \citet{2017ApJ...841..127M} recently determined the
distribution of nickel masses quite reliably from the tail phase based on a
sample whose representative character was established by comparison with the
LOSS survey \citep{2011MNRAS.412.1441L}, finding a median of $0.031\,\msun$ and
a mean of $0.046\,\msun$. For stripped-envelope SNe, the distribution of
explosion properties is less securely established. Although a number of studies
have investigated light curve parameters and/or explosion properties for larger
samples of stripped-envelope SNe \citep[e.g.][]{2011ApJ...741...97D,
2014ApJS..213...19B, 2016MNRAS.457..328L, 2016MNRAS.458.2973P,
2019MNRAS.485.1559P, 2018A&A...609A.136T} these are usually not based on
volume-limited surveys (for an exception, see \citealt{2011MNRAS.412.1441L}),
and nickel masses are often inferred from the peak luminosity using Arnett's
Rule \citep{1982ApJ...253..785A}.

Although the precise impact of selection biases for transients is not
straightforward to determine, the classical formula for the Malmquist bias
furnishes a rough estimate. Based on a standard deviation $0.71\texttt{-}0.78 \,
\mathrm{mag}$ in peak luminosity in the SN~Ib/c sample of
\citet{2016MNRAS.458.2973P}, the average peak luminosity may be overestimated by
about $0.7\texttt{-}0.8 \, \mathrm{mag}$ in samples that are not volume-limited,
which translates into a factor of two in nickel mass. The differences between
the volume-limited sample of \citet{2011MNRAS.412.1441L} and other population
studies of stripped-envelope explosions are indeed of similar magnitude.
Arnett's rule also tends to systematically overestimate nickel masses in Ib/c
SNe by about $50\%$ \citep{2015MNRAS.453.2189D}. It is therefore likely that the
average nickel mass in stripped-envelope explosions (excluding broad-lined Ic
SNe) is only $\mathord{\sim}70\%$ higher than for Type IIP and not by a factor
of five as suggested by \citet{2019A&A...628A...7A}.

Our models indicate that modestly higher nickel masses (about 50\%) in
stripped-envelope explosions might be explained naturally as a consequence of
binary mass transfer and its impact on the subsequent stages of stellar
evolution. After accounting for biases in observationally inferred nickel
masses, there may be no strong need to invoke magnetar powering for a
substantial fraction of stripped-envelope SNe as advocated by
\citet{2020ApJ...890...51E}. Observed outliers among Ib/c SNe with high peak
luminosity and nickel masses ${\gtrsim}\,0.20\texttt{-}0.25$
(Fig.~\ref{fig:Eexpl-vkick-MNi-vs-compactness}c) still cannot be directly
explained by our sample of stripped-envelope progenitors. However, viewing angle
effects could partly account for such events; moderate asymmetries in the nickel
distribution may well increase the peak luminosity and the ``apparent'' nickel
mass by about $0.5\, \mathrm{mag}$ or $60\%$ \citep{2007MNRAS.378....2S}.

The calibration of our kick velocities include single and stripped stars and is
made with respect to the observed space velocities of pulsars as found by
\citet{2005MNRAS.360..974H}, \ie to a Maxwellian distribution with
$\sigma=265\,\kms$. Here, we find that stripped stars receive an intrinsically
larger kick ($\sigma=315\pm24\,\kms$) than single stars
($\sigma=222\pm23\,\kms$). While these $\sigma$ values are not dramatically
different from the \citeauthor{2005MNRAS.360..974H} distribution (about $\pm
15\text{--}20\%$), they are, \eg, relevant to interpret the observed space
velocities of X-ray and other compact-object binaries, they would reduce rate
predictions of compact-object mergers from isolated binaries and slightly
increase expectations for space velocities of runaway/walkaway stars \citep[see
\eg][]{2019A&A...624A..66R}.

As stressed in Sect.~\ref{sec:parametric-sn-code}, we only consider mean kick
velocities and not also the dispersions that are inherent to stochastic kick
formation. While the diversity of our pre-SN models gives rise to a distribution
of mean kick velocities that appears to be generally compatible with
observations of, \eg, \citet{2005MNRAS.360..974H}, the stochastic nature of
kicks will further affect the shape of the distribution such that a more
in-depth comparison to observations cannot be made at this point.

%
%
\section{Conclusions}\label{sec:conclusions}

Most massive stars reside in binary systems such that they will exchange mass
with their companions at some point during their lives. Here, we study how
\casea, B and~C envelope stripping affects the further evolution of stars up to
core collapse and thereby their explosion properties. The stars are modelled
using the \mesa stellar evolution code and the SN stage is analysed with the
parametric SN model of \citet{2016MNRAS.460..742M}. Our main results can be
summarized as follows.

\begin{itemize}
    \item Because of the removal of the hydrogen-rich envelope, \casea and~B
    stripped stars have larger core carbon abundances $X_\mathrm{C}$ after core
    helium burning than single and \casec stripped stars for the same CO core
    masses $M_\mathrm{CO}$ -- the \casea and~B stripped stars form a distinct
    branch in the $M_\mathrm{CO}$--$X_\mathrm{C}$ plane
    (Fig.~\ref{fig:core-c-mass-fraction}).
    \item The two key quantities $M_\mathrm{CO}$ and $X_\mathrm{C}$ largely
    determine the core evolution through the advanced nuclear burning phases up
    to core collapse. Because of the distinct differences of stripped and single
    stars, we find different pre-SN structures in these stars. For example, the
    \casea and~B stripped stars have on average lower iron core masses for the
    same $M_\mathrm{CO}$ compared to single and \casec stripped stars, and the
    compactness parameter $\xi_{2.5}$ is also different. Both single and
    stripped stars show a compactness peak at certain $M_\mathrm{CO}$ values
    related to the transition from convective to radiative core carbon burning.
    In both cases, there is also a steep increase in $\xi_{2.5}$ beyond some
    $M_\mathrm{CO}$. Importantly, the compactness peak and steep increase in
    compactness are at higher $M_\mathrm{CO}$ in \casea and~B stripped stars
    than in single and \casec stripped stars.
    \item The structural differences translate into differences in the SN
    properties. Compared to single stars, \casea and~B stripped stars result in
    on average lower NS and BH masses ($\Delta M_\mathrm{NS} \approx 0.05\,
    \msun$), higher explosion energies ($\Delta E_\mathrm{expl} \approx 0.2\,
    \bethe$), larger kicks ($\Delta v_\mathrm{kick} \approx 100\, \kms$) and
    larger nickel yields ($\Delta M_\mathrm{Ni} \approx 0.02\, \msun$). Kick
    velocities of ${\gtrsim}\,600\,\kms$ are almost exclusively found in
    stripped stars. Setting BHs from stars in the compactness peak aside, single
    and \casec stripped stars give rise to BHs for initial masses
    $M_\mathrm{ini}\gtrsim35\,\msun$ while this limit shifts to
    $M_\mathrm{ini}\gtrsim 70\,\msun$ in \casea and~B stripped stars.
    \item Some models beyond the compactness peak likely experience significant
    fallback and probably a weak or failed SN. The BHs formed from envelope
    material falling back onto the NS receive a kick that is substantially lower
    than that of the original NS (depending on the amount of fallback). 
    \item The NS mass distribution is unimodal with a tail extending up to the
    maximum allowed NS mass of $2.0\,\msun$, \ie some NSs are born massive (we
    do not consider the contributions from ECSNe and ultra-stripped stars). The
    BH mass distribution is multimodal: stars falling into the compactness peak
    give rise to an island of BH formation, and the lowest mass BHs are always
    from stripped stars. The maximum BH mass from our stripped evelope models is
    about $20\,\msun$ (that of single stars is larger and depends on the
    amount of mass loss during LBV phases). There is a gap between NSs and BHs
    in our models if we discard fallback. The gap may (partly) disappear
    depending on the amount of fallback (low fallback fractions reduce the gap
    and high fallback fractions maximize it). The features in the compact
    remnant mass distributions give rise to distinct peaks in the chirp-mass
    distribution of NS-NS, BH-NS and BH-BH mergers.
    \item At least one star (if not both stars) in compact object mergers from
    isolated binary evolution will have evolved through envelope stripping.
    Because the initial mass range for NS formation is largely enhanced in
    stripped stars, we find a significant reduction of BHs in stellar
    populations, \eg the NS to BH ratio is ${\approx}\,15$ in our \caseb stripped
    models whereas it is only 2--3 for our stellar models when applying
    the delayed and rapid SN models of \citet{2012ApJ...749...91F} that are
    frequently applied in state-of-the-art population synthesis models.
    Consequently, the expected BH-NS and BH-BH merger rates are reduced and, from a simplified
    population synthesis model, we
    find reductions of 25--50\% and 90\%, respectively.
    \item Relative to the \citet{2012ApJ...749...91F} SN model, the NS-NS merger
    rate in our default population model increases only marginally (6--15\%).
    This is because a large fraction of NS-NS mergers forms from low-kick SNe
    (\eg ECSNe and stars with low iron cores) that are not affected by the
    increase in the initial-mass parameter space for NS formation from stripped
    stars.
\end{itemize}

We conclude that the removal of the hydrogen-rich envelope not only naturally
affects the SN type (II vs Ib/c), but has severe implications for the interior
structure and evolution of stars, and thus the SN explosion itself. This will in
turn help to further our understanding of NS and BH formation, the explosion
properties of SNe, compact object binaries and gravitational-wave merger events.

\begin{acknowledgements}
BM was supported by ARC Future Fellowship FT160100035.
We thank the referee for carefully reading our manuscript and the useful
comments that helped to improve it. Moreover, we thank Danny Vanbeveren for
pointing out that NS formation for stripped stars of initially up to $70\,\msun$
can explain why most of the observed WR+OB binaries may not produce wind-fed
(BH) high-mass X-ray binaries and thereby helps to understand a discrepancy they
reported in \citet{2020A&A...636A..99V}.
This research made use of NumPy \citep{oliphant2006numpy}, SciPy
\citep{2020NatMe..17..261V}, Matplotlib \citep{hunter2007matplotlib} and Jupyter
Notebooks \citep{kluyver2016jupyternotebook}.
\end{acknowledgements}

\appendix

\section{Relation of the parametric SN model to structural parameters of the pre-SN models}\label{sec:relation-SN-code-structure}

\begin{figure*}
    \begin{centering}
    \includegraphics{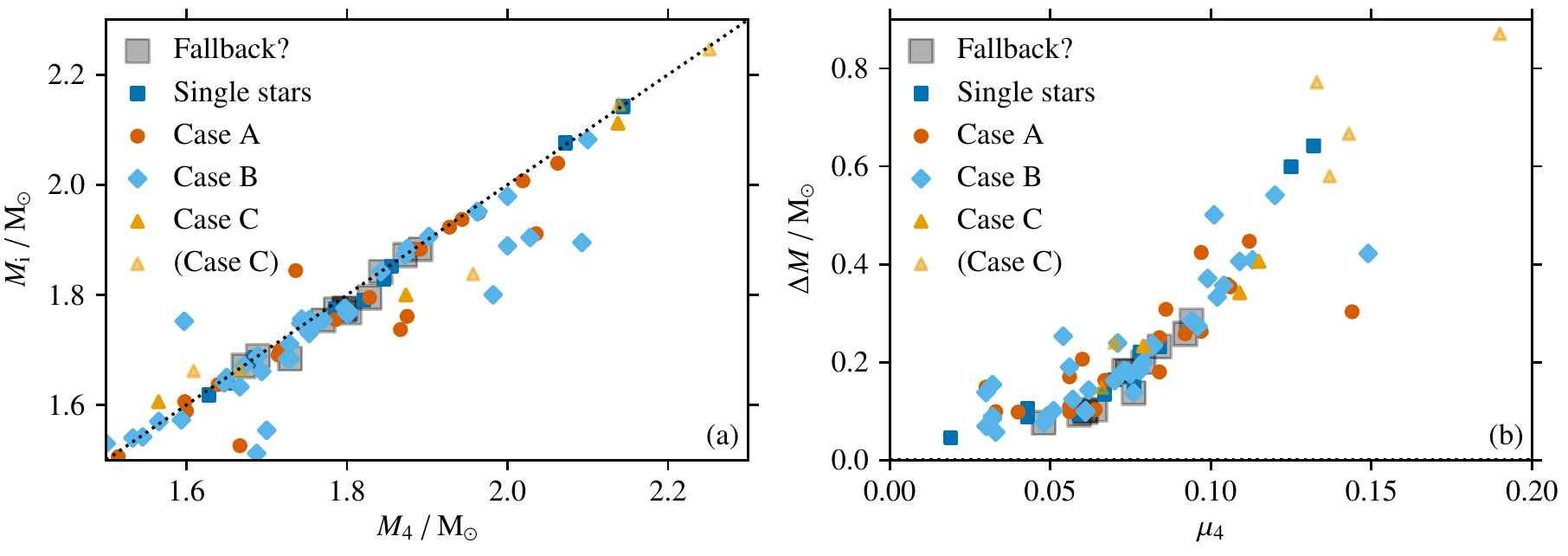} 
    \par\end{centering} \caption{Comparison of $M_\mathrm{i}$ and $\Delta M$ of
    the parametric SN model used in this work \citep{2016MNRAS.460..742M} with
    the structural parameters $M_4$ and $\mu_4$ of the pre-SN models
    \citep[\cf][]{2016ApJ...818..124E}. See text for details.}
    \label{fig:relation-SN-code-structure}
\end{figure*}

The mass coordinate at which shock revival occurs, $M_\mathrm{i}$, and the mass
accreted onto the proto-NS driving the neutrino luminosity, $\Delta M$, in the
applied parametric SN model \citep{2016MNRAS.460..742M} are closely related to
structural parameters of the pre-SN models used by \citet{2016ApJ...818..124E}
to classify the explodability of stars. \citet{2016ApJ...818..124E} define the
mass coordinate at which the dimensionless entropy per nucleon $s=4$,
\begin{equation}
    M_4 = m(s=4)/\msun,
    \label{eq:M4}
\end{equation}
and the radial mass derivative at this mass coordinate,
\begin{equation}
    \mu_4 = \left. \frac{\mathrm{d}m/\msun}{\mathrm{d}r/1000\,\mathrm{km}} \right\rvert_{s=4}.
    \label{eq:mu4}
\end{equation}
In physical terms, $M_4$ indicates the base of the oxygen shell (located at
$s\approx4\text{--}6$) and is used as a proxy of the mass of the proto-NS,
$M_\mathrm{NS}$, similar to the meaning of $M_\mathrm{i}$ in the parametric SN
model of \citet{2016MNRAS.460..742M} (see also
Sect.~\ref{sec:parametric-sn-code}). The spatial mass derivative at the base of
the O shell indicates the mass accretion rate $\dot{M}$ onto the proto-NS stars
\citep{2016ApJ...818..124E} and the product $\mu_4 M_4\propto M_\mathrm{NS}
\dot{M}$ is then a proxy of the neutrino luminosity, $L_\nu \propto G
M_\mathrm{NS} \dot{M}/R_\mathrm{NS}$, where $R_\mathrm{NS}$ is the radius of the
proto-NS. Hence, the mass accreted onto the proto-NS in the parametric SN model
$\Delta M$ is directly linked to $\mu_4$, $\Delta M \propto \dot{M}
\tau_\mathrm{acc}$ with $\tau_\mathrm{acc}$ the timescale over which the
proto-NS accretes mass and powers the SN explosion. Indeed, these
correspondences are also found quantitatively in our models as shown in
Fig.~\ref{fig:relation-SN-code-structure}. In particular, the mass coordinate of
shock revival is found to be closely related to the base of the O shell at
$s=4$.

\section{Fitting formulae}\label{sec:fit-functions}

\begin{figure*}
    \begin{centering}
    \includegraphics{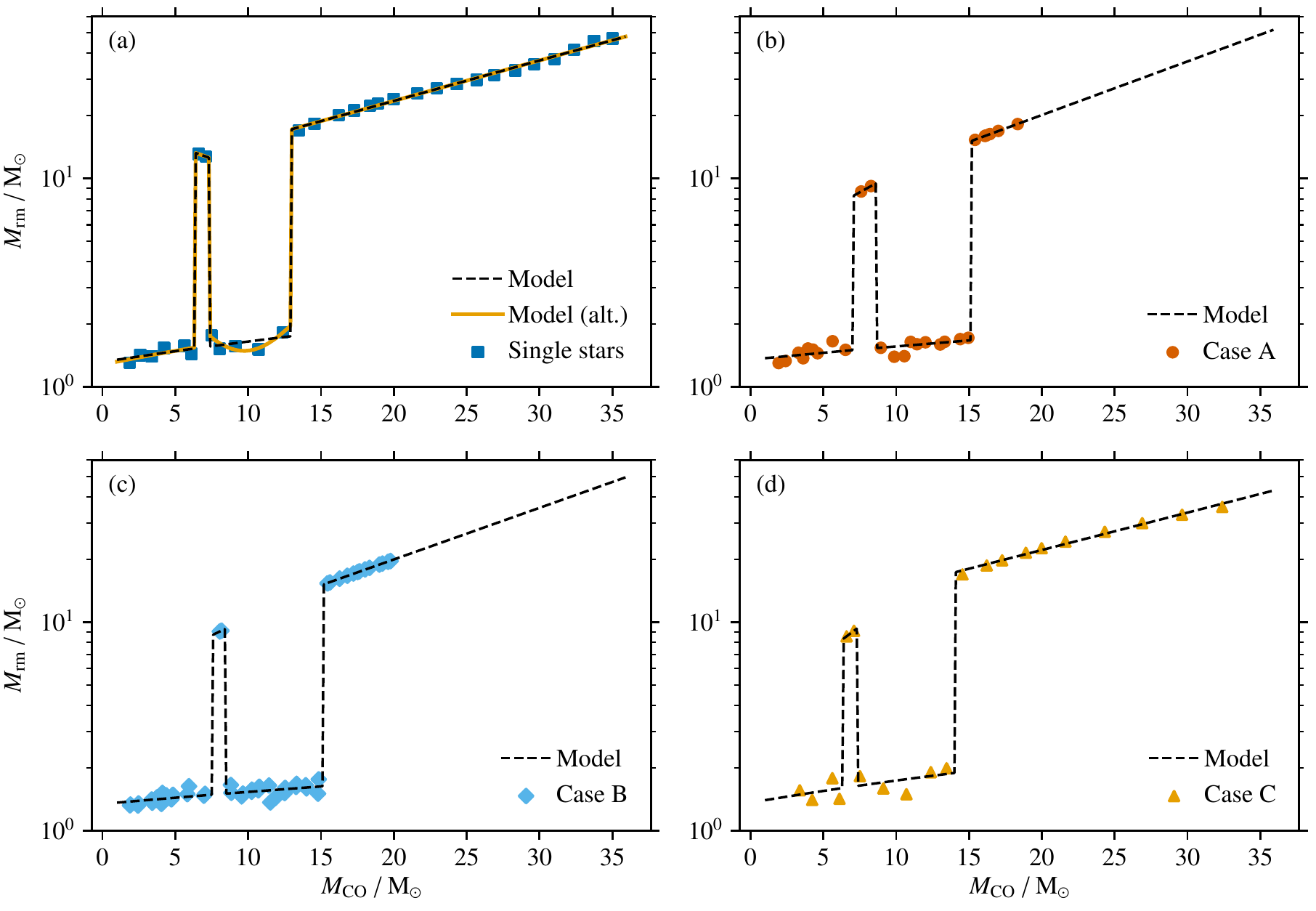} 
    \par\end{centering}
    \caption{Piecewise fits to the compact-object masses of single, \casea, B and~C stripped stars (panels a, b, c and d, respectively). For the single stars in panel (a), we also show an alternative model fit, ``Model (alt.)''.}
    \label{fig:com-toy-models}
\end{figure*}

We here summarise fitting formulae to the compact object masses of our single
and stripped binary star models (Eqs.~\ref{eq:single}--\ref{eq:case-c} below).
Some of these fits are used in the toy population models described in
Sect.~\ref{sec:pop-syn}. Below, we define $x=M_\mathrm{CO}/\msun$, \ie $x$ is
the CO core mass at the end of core helium burning. For the fitting functions,
we neglect fallback. In order to relate $M_\mathrm{CO}$ to initial masses, we
use spline fits as shown in Fig.~\ref{fig:mini-mco}. The fits to the compact
remnant masses are visualized in Fig.~\ref{fig:com-toy-models} and the fitting
functions are as follows.

\paragraph*{Single stars:}
\begin{align}
\log\,\frac{M_\mathrm{rm}}{\msun} = 
    \begin{cases}
        \log\,(0.03357\,x + 1.31780) & x < 6.357 \\
        -0.02466\,x + 1.28070 & 6.357 \leq x < 7.311 \\
        \log\,(0.03357\,x + 1.31780) & 7.311 \leq x < 12.925 \\
        0.01940\,x + 0.98462 & x \geq 12.925
    \end{cases}
    \label{eq:single}
\end{align}

\paragraph*{Single stars (alternative):} 
Employing a parabolic fit function to NS masses beyond the compactness peak
instead of the linear fit function used above.
\begin{align}
\log\,\frac{M_\mathrm{rm}}{\msun} = 
    \begin{cases}
        \log\,(0.04199\,x + 1.28128) & x < 6.357 \\
        -0.02466\,x + 1.28070 & 6.357 \leq x < 7.311 \\
        \log\,(0.04701\,x^2 - 0.91403\,x & \\ 
        \quad\quad + 5.93380) & 7.311 \leq x < 12.925 \\
        0.01940\,x + 0.98462 & x \geq 12.925
    \end{cases}
    \label{eq:single-alt}
\end{align}

\paragraph*{\casea:}
\begin{align}
\log\,\frac{M_\mathrm{rm}}{\msun} = 
    \begin{cases}
        \log\,(0.02128\,x + 1.35349) & x < 7.064 \\
        0.03866\,x + 0.64417 & 7.064 \leq x < 8.615 \\
        \log\,(0.02128\,x + 1.35349) & 8.615 \leq x < 15.187 \\
        0.02573\,x + 0.79027 & x \geq 15.187
    \end{cases}
    \label{eq:case-a}
\end{align}

\paragraph*{\caseb:}
\begin{align}
\log\,\frac{M_\mathrm{rm}}{\msun} = 
    \begin{cases}
        \log\,(0.01909\,x + 1.34529) & x < 7.548 \\
        0.03306\,x + 0.68978 & 7.548 \leq x < 8.491 \\
        \log\,(0.01909\,x + 1.34529) & 8.491 \leq x < 15.144 \\
        0.02477\,x + 0.80614 & x \geq 15.144
    \end{cases}
    \label{eq:case-b}
\end{align}

\paragraph*{\casec:}
\begin{align}
\log\,\frac{M_\mathrm{rm}}{\msun} = 
    \begin{cases}
        \log\,(0.03781\,x + 1.36363) & x < 6.357 \\
        0.05264\,x + 0.58531 & 6.357 \leq x < 7.311 \\
        \log\,(0.03781\,x + 1.36363) & 7.311 \leq x < 14.008 \\
        0.01795\,x + 0.98797 & x > 14.008
    \end{cases}
    \label{eq:case-c}
\end{align}

\section{Schematic mass distribution of compact objects}\label{sec:comf-toy-model}

\begin{figure*}
    \begin{centering}
    \includegraphics{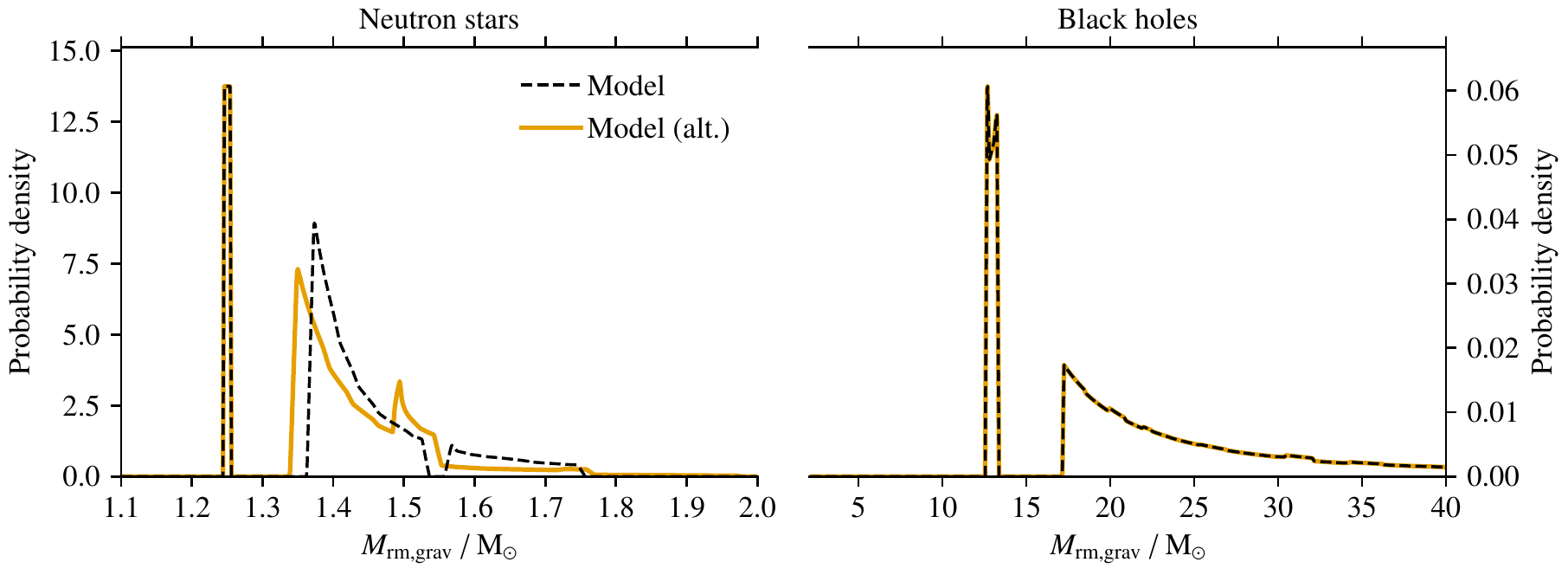} 
    \par\end{centering} \caption{Mass distribution of NSs and BHs of the two
    models from Fig.~\ref{fig:com-toy-models}a. Despite the seemingly small
    difference between the NS masses of the two models, the NS-mass distribution
    can show distinctly different features around masses that correspond to
    stars in the compactness peak. A contribution from ECSNe at $1.25\,\msun$
    has been added, but such SN progenitors are not studied here.}
    \label{fig:comf-toy-models}
\end{figure*}

The NS-mass distribution of single stars in
Fig.~\ref{fig:comf-default-and-fallback-model} shows a gap or dearth in the mass
range $1.6\text{--}1.7\,\msun$ that is caused by failed SNe from stars in the
compactness peak. The existence of such a gap also depends on the exact NS
masses of stars just around the compactness peak as we show here. We use two
different, piecewise fitting functions to the compact-remnant masses of single
stars (Fig.~\ref{fig:com-toy-models}). Model~1 assumes monotonically-increasing
NS masses with $M_\mathrm{CO}$, whereas we fit a parabolic function to the NS
masses beyond the compactness peak in Model~2. Assuming a power-law initial mass
function with high-mass slope $\gamma\,{=}\,-2.3$ as in
Sect.~\ref{sec:ns-bh-mass-distr}, we find the NS and BH mass distributions in
Fig.~\ref{fig:comf-toy-models}. 

The NS-mass distributions of Model~1 and~2 show distinctly different features
despite the seemingly small changes in the models. A gap is visible in Model~1
at $1.55\,\msun$ and is absent in Model~2. Instead of the gap, a bump is
observed at $1.50\,\msun$, which corresponds to the local minimum in the NS-mass
fit of Model~2 at $M_\mathrm{CO}{\approx}\,10\,\msun$. The shift at the lowest
NS masses between both models is caused by a slightly different fit to the NS
masses before the compactness peak that is hardly visible in
Fig.~\ref{fig:com-toy-models}.

\bibliographystyle{aa}

\end{document}